\documentclass[aps,pra,twocolumn,superscriptaddress,nofootinbib,showpacs]{revtex4-1}

\usepackage{lipsum}
\usepackage{braket}
\usepackage{graphicx}
\usepackage{dcolumn}
\usepackage{bm}
\usepackage{amsmath}
\usepackage{subfigure}

\newcommand{\ev}[1]{\langle #1 \rangle}
\newcommand{\lv}[0]{\mathcal{L}}
\newcommand{\pt}[0]{\mathcal{V}}
\newcommand{\td}[1]{\tilde{#1}}
\newcommand{\nbar}[0]{\bar{n}}
\newcommand{\lvket}[1]{\lvert #1 \rangle \rangle}
\newcommand{\lvbra}[1]{\langle \langle #1 \rvert}

\newcommand{\tn}[0]{\tau,\text{ph}}
\newcommand{\neff}[0]{\text{ph}}
\newcommand{\sigsub}[0]{\sigma}
\newcommand{\COM}[0]{\text{COM}}
\newcommand{\sn}[0]{\sigma,\text{ph}}

\newcommand{\subfigimg}[4][,]{%
  \setbox1=\hbox{\includegraphics[#1]{#4}}
  \leavevmode\rlap{\usebox1}
  \rlap{\hspace*{#3}\raisebox{\dimexpr\ht1-2\baselineskip}{#2}}
  \phantom{\usebox1}
}

\begin{document}

\title{Steady-state spin synchronization through the collective motion of 
trapped ions}

\author{Athreya Shankar}
\email[]{athreya.shankar@colorado.edu}
\affiliation{JILA, NIST, and Department of Physics, University of Colorado 
Boulder, Boulder, Colorado 80309-0440 USA}
\author{John Cooper} 
\affiliation{JILA, NIST, and Department of Physics, University of Colorado 
Boulder, Boulder, Colorado 80309-0440 USA}
\author{Justin G. Bohnet}
\affiliation{Time and Frequency Division, National Institute of Standards 
and Technology, Boulder, Colorado 80305, USA}
\author{John J. Bollinger}
\affiliation{Time and Frequency Division, National Institute of Standards 
and Technology, Boulder, Colorado 80305, USA}
\author{Murray Holland}
\affiliation{JILA, NIST, and Department of Physics, University of Colorado 
Boulder, Boulder, Colorado 80309-0440 USA}

\date{\today}

\begin{abstract}
Ultranarrow-linewidth atoms coupled to a lossy optical cavity mode 
synchronize, i.e. develop correlations, and exhibit steady-state 
superradiance when continuously repumped. This type of system displays 
rich collective physics and promises metrological applications. These 
features inspire us to investigate if analogous spin synchronization 
is possible in a different platform that is one of the most robust and 
controllable experimental testbeds currently available: ion-trap systems. 
We design a system with a primary and secondary species of ions that 
share a common set of normal modes of vibration. In analogy to the 
lossy optical mode, we propose to use a lossy normal mode, obtained 
by sympathetic cooling with the secondary species of ions, to mediate 
spin synchronization in the primary species of ions. Our numerical study 
shows that spin-spin correlations develop, leading to a macroscopic 
collective spin in steady-state. We propose an experimental method based 
on Ramsey interferometry to detect signatures of this collective spin; 
we predict that correlations prolong the visibility of Ramsey fringes, 
and that population statistics at the end of the Ramsey sequence can be 
used to directly infer spin-spin correlations.
\end{abstract}


\maketitle

\section{\label{sec:introduction}Introduction}

Steady-state synchronization of atomic dipoles forms the foundation for 
ultra-stable optical lasers utilizing narrow-linewidth atoms coupled to 
a lossy cavity mode. Such lasers have recently been proposed
\cite{meiserPRL2009, meiserPRA2010} and experimentally explored with a 
Raman system \cite{bohnetNat2012}, and in a true narrow-linewidth 
transition in strontium \cite{norciaPRX2016}.
The cavity mode acts as a channel 
for synchronization of the atomic dipoles (spins) resulting in a macroscopic 
collective dipole in steady-state composed of correlated atoms 
\cite{meiserPRA2010}. Synchronization here refers to the development 
of a preferred relative phase (correlations) between every pair of spins.  
The output light is a result of collective spontaneous emission 
of this macroscopic 
dipole, as in the case of Dicke superradiance \cite{dickePR1954}, 
with the difference that the superradiance is in steady-state with 
repumping of the atoms balancing the cavity loss. 

Steady-state superradiant lasers provide a platform for studying quantum 
synchronization and have applications as ultra-stable optical frequency 
sources. The linewidth of the output light is determined by the  
decay rate of the narrow-linewidth transition \cite{meiserPRL2009}, 
exploiting the all-to-all pair-wise phase-locking of 
a large number of spins to drastically reduce the linewidth.
The exciting features of cavity steady-state superradiance, 
such as the narrow linewidth light and the spin synchronization, 
motivate us to ask whether a superradiance model can be used to 
synchronize quantum ensembles in other platforms, and if such 
systems could exhibit interesting physics and have possible applications. 

Ion-trap systems have become a robust platform for experiments related to 
quantum computing, simulation and metrology 
\cite{haffnerPR2008, blattNat2012, brittonNat2012}, making them an excellent 
candidate for studies of spin synchronization. Ion traps have long 
trapping times, routinely trapping ions for several hours. The incoherent 
repumping, crucial to maintain steady-state superradiance, introduces 
recoil heating which can kick neutral atoms out of the shallow traps 
used in optical cavities. Complicated schemes must be used to mimic a 
steady-state number of atoms in this situation. However, this problem 
is negligible in ion traps which have much deeper trapping potentials. 
Further, ions in a trap are distinguishable because of the large 
spacings ($\sim \mu \text{m}$) between them, enabling
access to individual spins for direct measurement of spin-spin correlations.
  
One approach to synchronizing ions is to place 
ion traps in optical cavities, allowing the ions to interact with the cavity 
mode. However, the low density of trapped ions makes 
it difficult to couple more than $O(10^3)$ ions to the cavity, 
prohibiting the large collective cooperativities possible with neutral 
atoms, where $10^5$ to $10^6$ atoms are routinely used. 

A second approach is to couple ions through the normal modes of vibration 
of the trap, arising out of the Coulomb interactions between the ions. 
Like optical cavity modes, these normal modes are a natural coupling 
channel for interactions between distant particles. A normal mode 
of vibration and an optical cavity mode are both bosonic modes that 
can be described in the language of quantum harmonic oscillators 
\cite{[The analogy between a vibrational mode and an optical mode 
has been previously exploited to create a phonon laser with a 
single trapped ion; see ] vahalaNat2009}. 
Laser beams can be used to couple the 
electronic and motional degrees of freedom in different ways 
\cite{sorensenPRL1999,liebfriedRMP2003}. Ion traps 
also enable us to engineer a dedicated dissipative channel with tunable 
properties: a subset of ions can be used to sympathetically cool the entire 
crystal \cite{larsonPRL1986,kielpinskiPRA2000,barrettPRA2003}, 
removing phonons from the normal modes analogous to lossy mirrors 
removing photons from the cavity mode. The phonon loss rate and equilibrium 
phonon number (temperature) can be controlled by adjusting the power and 
detuning of the cooling laser.   

In this paper, we follow this second approach, to design and analyze a scheme 
for generating spin synchronization in an ion trap, 
by coupling a collection of continuously repumped ions with a heavily 
damped normal mode of vibration. This scheme offers several features 
that are quite novel in ion trap systems. Most protocols in ion traps 
use Hamiltonian interactions. However, the present approach promises 
to synchronize a mesoscopic ($20$ to $500$) number of ions using 
dissipation as a crucial ingredient. Our proposal is enabled
by recent demonstrations of control over hundreds of ions in Penning traps 
\cite{biercukQIC2009}, as well as improvements in radio frequency (RF) traps 
\cite{monroeSci2013,kumphNJP2011} that make it possible 
to control tens of ions in these traps. The key ingredients have also been 
implemented with a small number of ions in RF traps for preparing 
entangled states, demonstrating the feasibility of 
our scheme \cite{linNat2013}.

Spin synchronization from steady-state superradiance can enhance metrology 
with trapped ions. Theoretical studies have shown that when continuously 
repumped spins interact with a heavily damped cavity mode during the 
interrogation time of a Ramsey pulse sequence, the resulting Ramsey 
fringes can decay at a rate much slower than the decay and dephasing 
rates for unsynchronized atoms \cite{xuPRL2015}. Implementing such a 
protocol using a damped normal mode in an ion trap 
could mitigate inhomogeneous broadening effects, and improve the capability 
of trapped ions for sensing, for example, of time-varying magnetic fields.

This paper is organized as follows. In Sec.~\ref{sec:model}, we 
consider a model of two species of ions loaded in an ion trap that can be 
used to explore spin synchronization mediated by a damped normal mode. In 
Sec.~\ref{sec:a_model_system}, we consider a specific example of an ion 
trap system where this scheme could be implemented. We numerically 
investigate this model system, comparing the results with the corresponding 
atom-cavity model. We look for signatures of synchronization brought about 
by steady-state superradiance such as the pair-wise correlations between 
ions. We also propose an experimental scheme to observe features of the 
collective dipole based on a Ramsey pulse sequence.
We show that the collective dipole results in Ramsey fringes that 
decay with a slower rate than that expected from incoherent repumping, 
and the variance in the population readout at the end of the Ramsey 
sequence directly measures the steady-state spin-spin correlations.
We then briefly touch on how this model can be a potential candidate for 
improving metrology with ion trap systems. We conclude by summarizing our 
results in Sec.~\ref{sec:conclusion}, and indicating possible future 
directions.

\section{\label{sec:model}Model}

\begin{figure*}[!htb]
\centering
\includegraphics[width=\linewidth]
{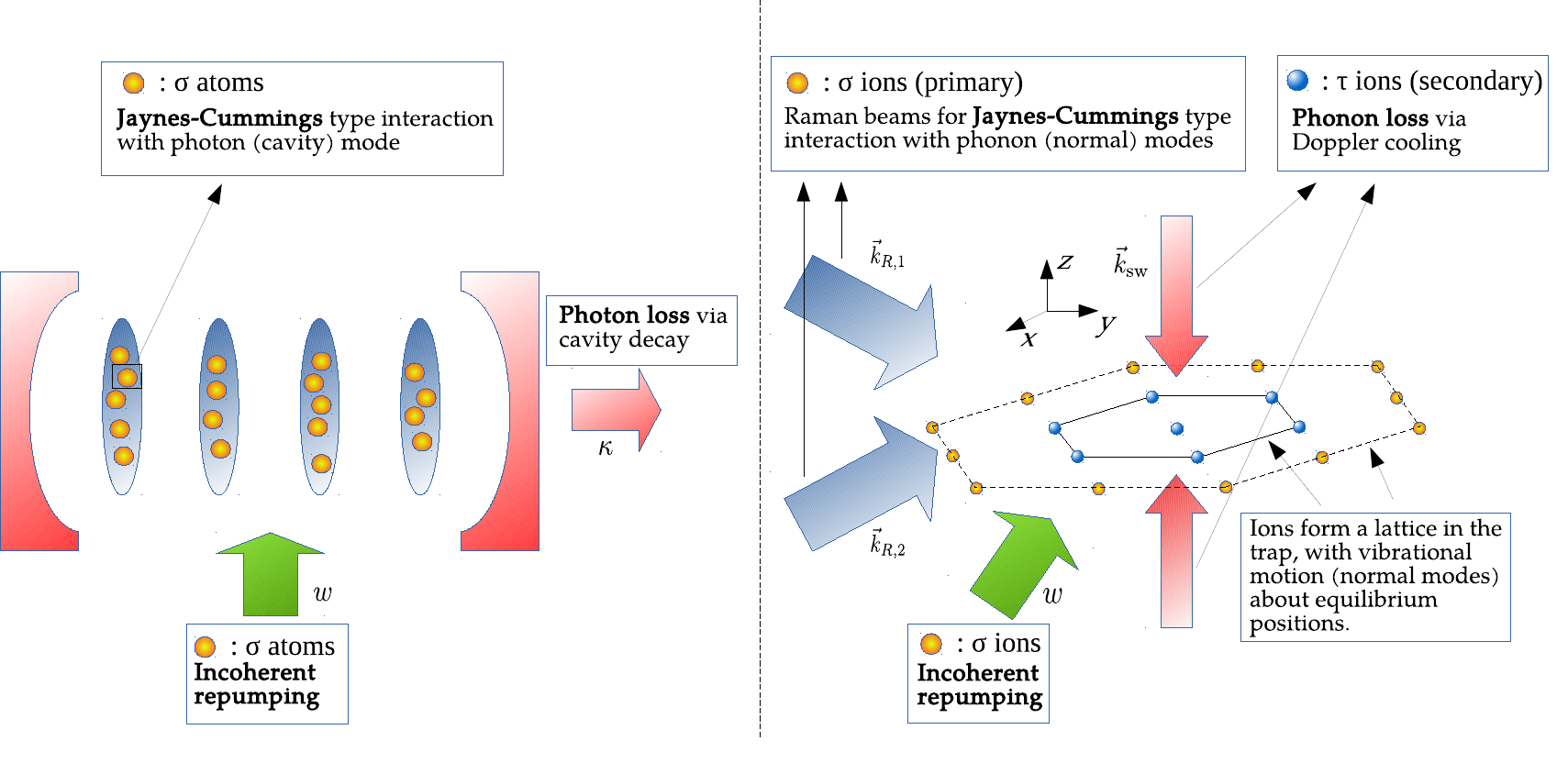}
\caption{(color online) Mapping cavity steady-state superradiance onto an ion trap system. 
In the left panel, we show the model for cavity steady-state superradiance, 
where the cavity mode serves as a mediator for collective decay of the spins 
formed by the $\sigma$ atoms. In the right panel, we show the ion trap 
system where a normal mode of vibration serves as a mediator for collective 
decay of the spins formed by the $\sigma$ ions. The figure illustrates the 
model with a 2-dimensional crystal.  More generally our model can also be 
applied to 1D crystals of ions.}
\label{fig:ion_trap_mapping}
\end{figure*}

There are three crucial ingredients to generate steady-state superradiance 
in a cavity (see Fig.~\ref{fig:ion_trap_mapping}): (a) a heavily damped 
cavity mode, (b) a Jaynes-Cummings interaction between two-level atoms and 
the nearly-resonant cavity mode, and (c) incoherent repumping of the 
two-level atoms to maintain steady-state. 

In Fig.~\ref{fig:ion_trap_mapping}, we schematically show the mapping of 
the problem of cavity steady-state superradiance onto an ion trap system. We 
consider two species of ions, $\tau$ (secondary) and $\sigma$ (primary), 
loaded in an ion trap \cite{bermudezPRL2013}. The two species could be, for 
example, two different elements, or isotopes of the same element. 
The system has a total of $N = N_\tau + N_\sigma$ ions, and 
therefore the transverse ($z$-axis) motion of any ion can be described using 
the $N$ transverse normal modes of the system. The $\tau$ ions 
are used to sympathetically cool the normal modes of vibration of 
the system of ions. The $\sigma$ ions provide the effective spins that 
synchronize through the interaction with a damped normal mode. 

In Sec.~\ref{sec:tau_ions}, we demonstrate that Doppler cooling of 
the two-level $\tau$ ions leads to an effective damping of the normal modes. 
The effective dynamics for each mode can be described as an interaction of a 
single-mode harmonic oscillator with a reservoir at a finite temperature. 
Then, in Sec.~\ref{sec:sigma_ions}, we derive the interaction of the 
three-level $\sigma$ ions with a pair of off-resonant Raman beams, 
taking into account the effects of 
dissipative processes. When the difference frequency of the 
Raman beams is suitably tuned, this 
interaction models a Jaynes-Cummings type interaction between an effective 
spin-$1/2$ system and a particular normal mode. Finally, in 
Sec.~\ref{sec:sigma_eff_spin_spin}, we consider the interaction between the 
spin-$1/2$ systems formed by the $\sigma$ ions and the strongly damped 
normal modes. We obtain an effective dynamics for these spin-$1/2$ systems, 
that consists only of spin-spin interactions. We then compare our ion trap 
model with the model for cavity steady-state 
superradiance \cite{meiserPRA2010_2}, and highlight 
the similarities in the dynamics, as well as the differences.  

\subsection{\label{sec:tau_ions}Doppler cooling of $\tau$ ions}

\begin{figure}[!htb]
\centering
\includegraphics[scale=0.5]{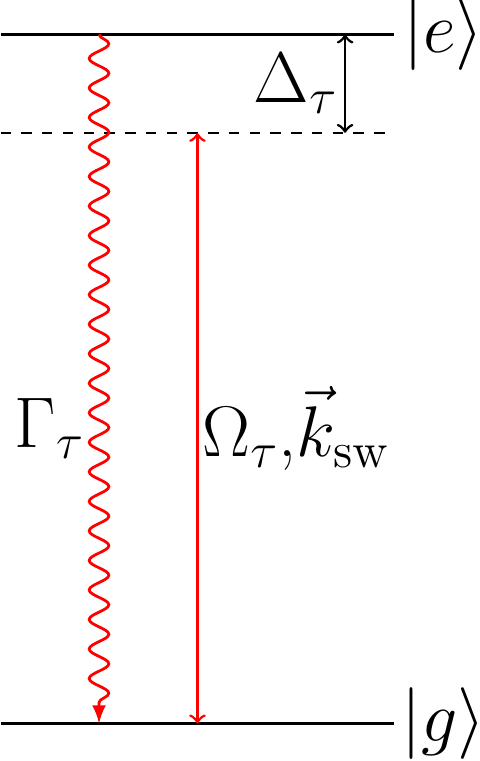}
\caption{(color online) Level diagram of a $\tau$ ion. The $\tau$ ions are driven using 
a cooling laser that is red detuned from the dipole allowed 
$\ket{e} \leftrightarrow \ket{g}$ transition. This results in cooling of the 
normal modes of vibration of the ion trap system.} 
\label{fig:tau_ion_level}
\end{figure}

The $\tau$ ions are two-level systems that are placed at the node of a 
standing-wave cooling laser \cite{ciracPRA1992}. A traveling-wave laser 
may be used for cooling provided the achieved steady-state temperature, 
characterized by the mean occupation number of the normal modes, 
is not very high. The level diagram of a $\tau$ ion is shown in 
Fig.~\ref{fig:tau_ion_level}. The $\ket{e} \leftrightarrow \ket{g}$ 
transition is dipole allowed, and can be used to Doppler cool the normal 
modes of the system. The level $\ket{e}$ decays to $\ket{g}$ at a rate 
$\Gamma_\tau$. The cooling laser has a Rabi frequency of $\Omega_\tau$ 
and a wavevector $\vec{k}_{\text{sw}} = k_{\text{sw}} \hat{z}$. We use the 
notation $\tau^\pm, \tau^z$ to denote the Pauli spin matrices associated 
with the $\tau$ ions. 

The master equation for the interaction of the $N_\tau$ $\tau$ ions and $N$ 
normal modes with the cooling laser is

\begin{eqnarray}
&&\dot{\rho}_{\tn} = -i[H_{\tn},\rho_{\tn}] \nonumber\\* 
&&\hphantom{\dot{\rho}_{\tn} =} + \frac{\Gamma_\tau}{2} 
\sum_m \int_{-1}^1 du W(u) \mathcal{D}[\tau_m^- e^{(-i k_\tau z_m u)}] 
\rho_{\tn}.\nonumber\\*
\label{eqn:tau_phonon_ME}
\end{eqnarray} 

Here, $\rho_{\tn}$ is the density matrix describing the $\tau$-spins and the 
normal modes (the subscript ``$\text{ph}$" is shorthand for ``phonons"). 
Throughout this paper, we have set $\hbar = 1$, 
unless we explicitly specify otherwise. The notation 
$\mathcal{D}[O]$ is used to represent the standard Lindblad dissipator, i.e., 
$\mathcal{D}[O] \rho = 2 O \rho O^\dagger 
- O^\dagger O \rho - \rho O^\dagger O$. 
The second term on the RHS of Eq.~(\ref{eqn:tau_phonon_ME}) accounts for 
the dissipation due to spontaneous emission, and its effects on the 
transverse motion of the ions. The wavevector $\vec{k}_\tau$ of the 
spontaneously emitted photon makes an angle $\theta = cos^{-1}u$ with 
the z-axis, where the distribution of the angles is given by the normalized, 
even function $W(u)$. The transverse position of the ion $m$ is denoted 
by $z_m$.

In a frame rotating at the cooling laser frequency, the Hamiltonian $H_{\tn}$ 
in Eq. (\ref{eqn:tau_phonon_ME}) is

\begin{eqnarray}
&&H_{\tn} = -\frac{1}{2} \Delta_\tau \sum_m \tau_m^z + 
\sum_n \omega_n b_n^\dagger b_n \nonumber\\*
&&\hphantom{H_{\tn} =} +\frac{\Omega_\tau}{2} 
\sum_m  \sin(k_{\text{sw}} z_m) (\tau_m^- + \tau_m^+),
\end{eqnarray}

\noindent where $\Delta_\tau = \omega_{\text{sw}} - (\omega_e - \omega_g)$ 
is the detuning of the cooling laser. The frequency of the normal mode $n$ 
is given by $\omega_n$, and its annihilation and creation operators are 
$b_n$ and $b_n^\dagger$.

For small detunings, $k_{\text{sw}} \approx k_\tau \equiv k$. The 
dimensionless quantity $k z_m$ for the ion $m$ can be expressed in terms of 
the normal modes of the system as

\begin{equation}
k z_m = \sum_n \eta_n^\tau \mathcal{M}_{mn} (b_n + b_n^\dagger),
\end{equation}

\noindent and captures the spread in the position of the ion relative to the 
wavelength of the light it interacts with. The quantity 
$\eta_n^\tau = k \sqrt \frac{\hbar}{2 m_\tau \omega_n}$ is the Lamb-Dicke 
parameter \cite{winelandJRNIST1998} for the normal mode $n$. 
The equilibrium positions of the $\sigma$ and $\tau$ ions
are due to a balance between the trap potential and the Coulomb interactions 
between the ions. Displacement of an ion from equilibrium results in simple
harmonic motion. The matrix $\mathcal{M}$ diagonalizes the potential energy
matrix (written in mass-weighted coordinates) of this simple harmonic motion.
The frequencies $\omega_n$ of the normal modes are obtained from the 
eigenvalues of this potential energy matrix 
\cite{[The analysis follows the classical treatment of normal modes discussed 
for eg. in ] goldstein2002classical}.

In the Lamb-Dicke regime ($\langle (k z_m)^2 \rangle^{1/2} \ll 1$) 
\cite{winelandJRNIST1998}, we can expand 
the RHS of the master equation in powers of \{$\eta_n^\tau$\}. When the 
decay rate $\Gamma_\tau$ is large compared with the couplings 
\{$\Omega_\tau \eta_n^\tau$\} between the system of normal modes and the 
reservoir of $\tau$ ions, second-order perurbation theory and
a Markov approximation can be used to 
arrive at an effective master equation for the damping of the system of 
normal modes (see Appendix~\ref{app:normal_mode_damping}). 
The cooling introduces couplings between the normal modes, resulting in a 
new dressed set of normal modes that are decoupled from each other. For 
simplicity, here we neglect couplings between different modes, and 
approximate the bare modes to be decoupled from each other\footnote{See 
Eq.~(\ref{eqn:doppler_full}) and the subsequent remarks in 
Appendix~\ref{app:normal_mode_damping}.}. Then, the effective master 
equation that describes the damping of normal modes is given by 

\begin{eqnarray}
&&\dot{\mu}_{\neff} = -i \left[ \sum_n \omega_n^\prime b_n^\dagger b_n, 
\mu_{\neff} \right] \nonumber\\*
&&\hphantom{\dot{\mu}_{\neff} =} + \sum_n D_n^- \mathcal{D}[b_n] \mu_{\neff} 
+ \sum_n D_n^+ \mathcal{D}[b_n^\dagger] \mu_{\neff},
\label{eqn:tau_phonon_final_ME}
\end{eqnarray}

\noindent where 

\begin{eqnarray}
&&\omega_n^\prime = \omega_n + R_n^- (\Delta_\tau + \omega_n) 
+ R_n^+ (\Delta_\tau - \omega_n), \; \text{and} \nonumber\\*
&& D_n^\pm = R_n^\pm \frac{\Gamma_\tau}{2} \quad \text{with} \nonumber\\*
&& R_n^\pm = 
\frac{ \sum_m (\frac{1}{2} \Omega_\tau \eta_n^\tau \mathcal{M}_{mn})^2 } 
{\frac{\Gamma_\tau^2}{4} + (\Delta_\tau \mp \omega_n)^2 }.
\end{eqnarray}

Here $\mu_{\neff}$ is the density matrix describing the normal modes. To draw 
an  analogy with cavity QED models, it is useful to define a cooling rate 
per mode $\kappa_n = 2 (D_n^--D_n^+)$ and a mean occupation number per mode 
$\bar{n}_n = D_n^+/(D_n^--D_n^+)$. Then Eq.~(\ref{eqn:tau_phonon_final_ME}) 
can be written as

\begin{eqnarray}
&&\dot{\mu}_\neff = -i \left[ \sum_n \omega_n^\prime b_n^\dagger b_n, 
\mu_\neff \right] \nonumber\\*
&&\hphantom{\dot{\mu}_\neff} 
+ \sum_n \frac{\kappa_n (\bar{n}_n+1)}{2} \mathcal{D}[b_n] \mu_\neff 
+ \sum_n  \frac{\kappa_n \bar{n}_n}{2} \mathcal{D}[b_n^\dagger] \mu_\neff.
\nonumber\\*
\label{eqn:mode_damping_cQED}
\end{eqnarray}

Eq.~(\ref{eqn:mode_damping_cQED}) describes the decay of $N$ individual 
harmonic oscillators with frequencies $\{\omega_n\}$, each respectively in 
contact with a reservoir in a thermal state with mean occupation number 
$\bar{n}_n$, at rates $\kappa_n$ \cite{carmichael1999statistical}.

\subsection{\label{sec:sigma_ions}Interaction of $\sigma$ ions with Raman 
beams}

\begin{figure}[!htb]
\centering
\includegraphics[scale=0.5]{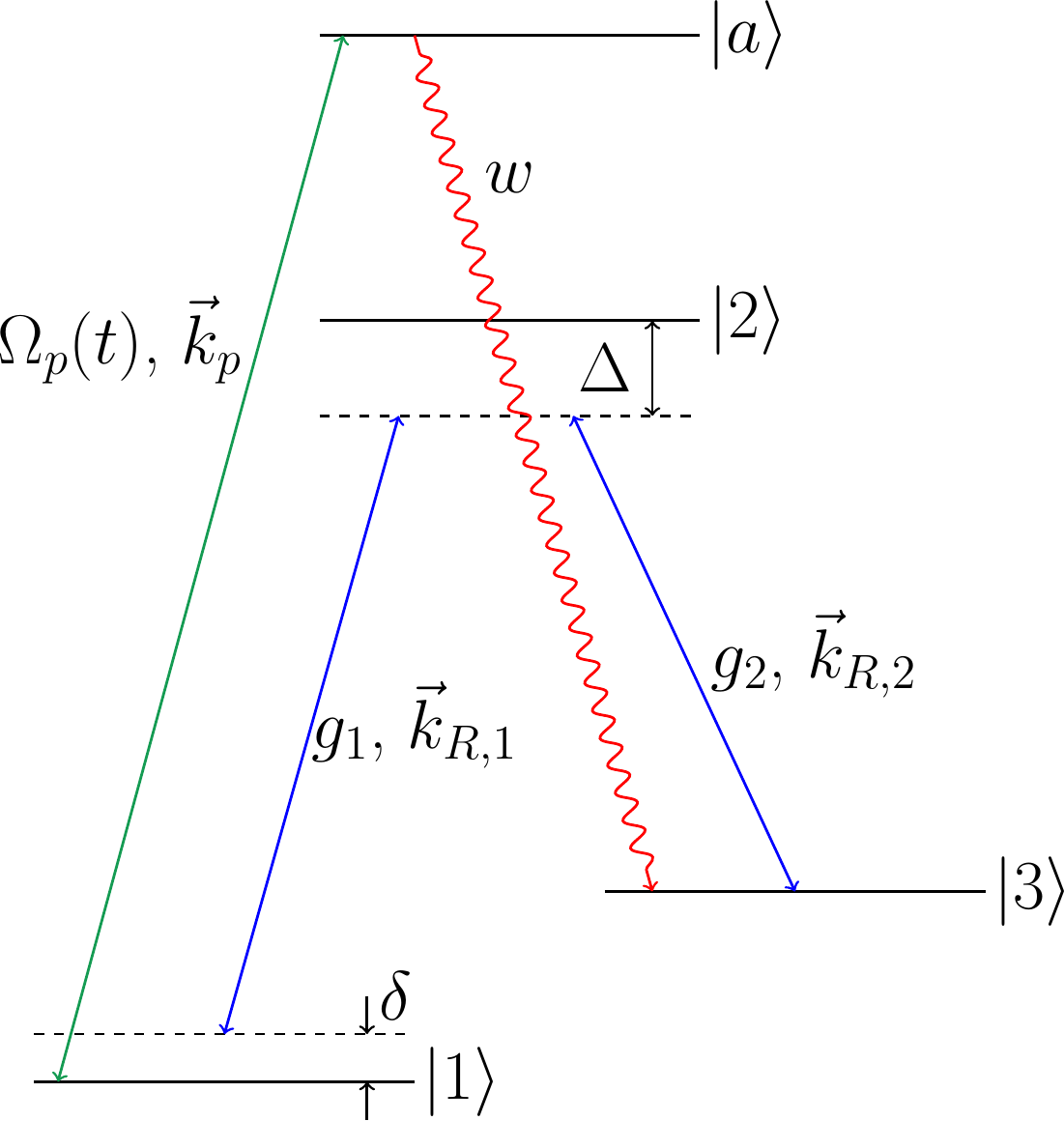}
\caption{(color online) Level diagram of a $\sigma$ ion. The three level configuration 
$\{ \ket{1}, \ket{2}, \ket{3} \}$ is used to drive stimulated Raman 
transitions in a far detuned regime, giving rise to an effective two-level 
system in the $\{ \ket{1}, \ket{3} \}$ manifold. The detuning $\delta$ can 
be adjusted to drive red sideband transitions coupling the electronic 
dynamics with an external normal mode of vibration. Incoherent pumping 
through an excited state $\ket{a}$ replenishes energy lost via Doppler 
cooling (not shown here) of the normal mode. } 
\label{fig:sigma_ion_level}
\end{figure}

The level diagram of a $\sigma$ ion is shown in 
Fig.~\ref{fig:sigma_ion_level}. The $\ket{1}\leftrightarrow\ket{2}$ and 
$\ket{3}\leftrightarrow\ket{2}$ transitions are dipole allowed, but the 
$\ket{1}\leftrightarrow\ket{3}$ transition is dipole forbidden. A pair of 
Raman beams are used to drive the $\ket{1}\leftrightarrow\ket{2}$ and 
$\ket{3}\leftrightarrow\ket{2}$ transitions. Their wavevectors, frequencies 
and Rabi coupling strengths are respectively 
$\vec{k}_{R,1}, \omega_{R,1}, g_1$ and $\vec{k}_{R,2}, \omega_{R,2}, g_2$.
The Rabi coupling strengths have a position dependency arising from the 
traveling-wave Raman beams, i.e.

\begin{equation}
g_1 = g_{1,0} e^{i\vec{k}_{R,1}\cdot\vec{x}}, \; \text{and} \; 
g_2 = g_{2,0} e^{i\vec{k}_{R,2}\cdot\vec{x}}.
\end{equation}  

The difference wavevector $\vec{k}_\sigma = \vec{k}_{R,1} - \vec{k}_{R,2}$ 
is along the (transverse) $z$-axis. The level $\ket{2}$ decays to levels 
$\ket{1}$ and $\ket{3}$ at rates $\Gamma_1$ and $\Gamma_2$ respectively. 
The Raman beams operate in a regime where they are far detuned from the 
transitions they drive: $\Delta_{1} = \omega_{R,1}-(\omega_2-\omega_{1}), 
\Delta_{2} = \omega_{R,2}-(\omega_2-\omega_{3})  \gg |g_1|, |g_2|, 
\Gamma_1, \Gamma_2$. 

The master equation for a $\sigma$-ion interacting with Raman beams is given 
by

\begin{equation}
\dot{\rho}_{\sigsub} = -i[H_{\sigsub},\rho_{\sigsub}] 
+ \frac{\Gamma_1}{2} \mathcal{D}[\sigma_{12}]\rho_{\sigsub} 
+ \frac{\Gamma_2}{2} \mathcal{D}[\sigma_{32}]\rho_{\sigsub},  
\label{eqn:sigma_Raman_ME}
\end{equation}    

\noindent where $\rho_{\sigsub}$ is the density matrix for a single 
$\sigma$-ion. 

The Hamiltonian appearing in Eq.~(\ref{eqn:sigma_Raman_ME}) is 

\begin{equation}
H_{\sigsub} = \Delta_1 \sigma_{11} + \Delta_2 \sigma_{33} + 
(\frac{g_1}{2} \sigma_{21} + \frac{g_2}{2} \sigma_{23} + \text{h.c.}),
\end{equation}

\noindent where we use the notation $\sigma_{ij} = \ket{i}\bra{j}, 
i,j = 1,2,3$ to represent operators acting on the electronic levels 
of the $\sigma$ ion. 

Driving this three-level system in a far detuned regime results in Rabi 
oscillations between levels $\ket{1}$ and $\ket{3}$. While this is a well 
known result \cite{winelandJRNIST1998}, 
it is important for our study to consider the dissipative 
processes that arise because of the scattering from $\ket{2}$. We use a 
recently developed Schrieffer-Wolff formalism for dissipative systems 
\cite{kesslerPRA2012}, which
is a projection operator method, to 
rigorously obtain the effective dynamics of the two-level system formed by 
the \{$\ket{1}$, $\ket{3}$\} manifold.  

The use of this formalism in the present case is detailed in 
Appendix~\ref{app:schrieffer_wolff}. We then get a description of the 
effective dynamics in the \{$\ket{1}$, $\ket{3}$\} manifold. We denote 
operators in this space using Pauli spin matrices: 
$\sigma^z = \sigma_{33}-\sigma_{11}, \sigma^+ = \sigma_{31}, 
\sigma^- = \sigma_{13}$. For a collection of 
$\sigma$ ions, the master equation describing the dynamics in the 
\{$\ket{1}$, $\ket{3}$\} manifold of these ions is then

\begin{eqnarray}
&& \dot{\mu} = -i[H^{\text{eff}},\mu] 
+ \frac{\Gamma_{31}}{2} \sum_l \mathcal{D}[\sigma_l^-]\mu 
\nonumber\\*
&& \hphantom{\dot{\mu} =}  
+ \frac{\Gamma_{13}}{2} \sum_l \mathcal{D}[\sigma_l^+]\mu 
+ \frac{\Gamma_{d}}{8} \sum_l \mathcal{D}[\sigma_l^z]\mu, 
\label{eqn:sigma_Raman_eff_ME}
\end{eqnarray}

\noindent where

\begin{equation}
H^{\text{eff}} = -\frac{1}{2}{\delta_R} \sum_l \sigma_l^z 
+ \sum_l \left (\frac{\Omega_{R,l}(z_l)}{2}\sigma_l^+ + \text{h.c.} \right). 
\end{equation}

Here, $\mu$ is the density matrix for the effective spin-$1/2$ systems 
formed by the $\ket{1},\ket{3}$ manifolds of the $\sigma$ ions. In writing 
Eq.~(\ref{eqn:sigma_Raman_eff_ME}), we have omitted certain 
`cross-terms' \cite{klimovOC2004}
 which eventually contribute at order 
$\Gamma_{1,2}^2/\Delta^2 (\ll 1)$ lesser than the interactions of interest. 
To avoid digressing, we outline the reasoning behind this 
omission in Appendix~\ref{app:schrieffer_wolff}. We have introduced several 
new symbols in Eq.~(\ref{eqn:sigma_Raman_eff_ME}), which are explained in 
Table~\ref{tab:sigma_Raman_eff_symbols}.

\squeezetable
\begin{table}
\caption{\label{tab:sigma_Raman_eff_symbols}Symbols used in writing the 
effective master equation for the $\sigma$ ions 
(Eq.~(\ref{eqn:sigma_Raman_eff_ME}))}
\begin{ruledtabular}
\begin{tabular}{ccc}
Symbol	&	Description	&	Expression \\ \hline \\ 
$\delta_R$	&	effective detuning	&	
$(\Delta_1 + \frac{|g_1|^2}{4\Delta_1}) - 
(\Delta_2 + \frac{|g_2|^2}{4\Delta_2})$ \\[5pt]
$\Omega_R$	&	effective Rabi frequency	&	
$\frac{g_1 g_2^*}{4} \left( \frac{1}{\Delta_1} 
+ \frac{1}{\Delta_2} \right)$ \\[5pt]
$\Delta$	&	average detuning	&	$
\frac{\Delta_1+\Delta_2}{2}$ \\[5pt]
$\Gamma_{31}$	&	effective spontaneous emission	&	
$\Gamma_1 \frac{|g_2|^2}{4 \Delta^2}$ \\[5pt]
$\Gamma_{13}$	&	effective incoherent repumping	&	
$\Gamma_2 \frac{|g_1|^2}{4 \Delta^2}$ \\[5pt]
$\Gamma_{d}$	&	effective dephasing	&	
$\Gamma_1 \frac{|g_1|^2}{4 \Delta^2} + \Gamma_2 \frac{|g_2|^2}{4 \Delta^2}$
\end{tabular}
\end{ruledtabular}
\end{table}

At this point, we consider a collection of $N_\sigma$ $\sigma$ ions and 
$N_\tau$ $\tau$ ions loaded in an ion trap. The collection of ions has 
$N = N_\sigma + N_\tau$ normal modes in total. The $\sigma$ ions with index 
$l, l \in \{1, \ldots, N_\sigma\}$ have an effective Rabi frequency 
$\Omega_{R,l}(z_l) = \Omega_{R}^0 e^{i k_\sigma z_l}$.  Once again, the 
dimensionless quantity $k_\sigma z_l$ for ion $l$ can be expressed in terms 
of the normal modes of the system as

\begin{equation}
k_\sigma z_l = \sum_n \eta_n^\sigma \mathcal{M}_{ln} (b_n + b_n^\dagger),
\end{equation}

\noindent where the quantity 
$\eta_n^\sigma = k_\sigma \sqrt \frac{\hbar}{2 m_\sigma \omega_n}$ is the 
Lamb-Dicke parameter for the normal mode $n$.

In the Lamb-Dicke regime ($\langle (k_\sigma z_l)^2 \rangle^{1/2} \ll 1$), 
the effective Rabi frequency $\Omega_{R,l}$ can be expanded up to first order 
as

\begin{equation}
\Omega_{R,l}(z_l) \approx \Omega_{R}^0  
+ i \Omega_{R}^0 \sum_n \eta_n^\sigma \mathcal{M}_{ln} (b_n + b_n^\dagger).
\end{equation}

The Raman lasers are now tuned to the red sideband 
\cite{winelandJRNIST1998} by adjusting the 
effective detuning $\delta_R$. If $|\Omega_R^0| \ll |\delta_R| \sim \omega_n$, 
the contributions from the carrier and blue sideband interactions can 
be neglected, as the coherences associated with these processes are 
$O(\Omega_R^0/\delta_R)$ and $O(\Omega_R^0 \eta_n^\sigma/\delta_R)$ 
respectively. The effective Hamiltonian in Eq.~(\ref{eqn:sigma_Raman_eff_ME}) 
is approximately

\begin{eqnarray}
&&H^{\text{eff}} \approx -\frac{1}{2}{\delta_R} \sum_l \sigma_l^z 
+ \sum_n \omega_n b_n^\dagger b_n \nonumber\\* 
&&\hphantom{H^{\text{eff}} \approx} + 
\sum_{l} \sum_{n} \left( \mathcal{F}_{ln} \sigma_l^+ b_n + \text{h.c.} 
\right),  
\end{eqnarray}

\noindent where 
$\mathcal{F}_{ln} = i \Omega_R^0 \eta_n^\sigma \mathcal{M}_{ln}/2$ is 
the effective coupling strength for a Jaynes-Cummings (JC) type interaction 
between ion $l$ and normal mode $n$. We have included the self-energy terms 
for the normal modes, since the master equation 
(\ref{eqn:sigma_Raman_eff_ME}) now describes the combined system of 
$\sigma$ ions and the normal modes. 
Note that $\mathcal{F}_{ln}$ is smaller than the effective Rabi frequency 
$\Omega_R^0$ by a factor $\eta_n^\sigma \mathcal{M}_{ln}$. This is the 
reason why the (usually) small dissipative processes arising from the 
stimulated Raman process could be important in our study.

\subsection{\label{sec:sigma_eff_spin_spin}Effective spin-spin model for 
$\sigma$ ions}

In Sec.~\ref{sec:sigma_ions}, we obtained the effective dynamics for the 
interaction of the $\sigma$ ions with the Raman lasers. The $\sigma$ ions, 
henceforth treated as effective spin-$1/2$ systems, interact with the normal 
modes through a Jaynes-Cummings type interaction. Earlier, in 
Sec.~\ref{sec:tau_ions}, the Doppler cooling of the $\tau$ ions was used to 
derive an effective damping for the normal modes. In this section, we 
proceed by describing the interaction of the $\sigma$ ions with 
these damped set of normal modes. 

The master equation for the interaction of the $\sigma$-ions with the damped 
set of normal modes is given by

\begin{eqnarray}
&&\dot{\rho}_{\sn} = -i [H_{\sn},\rho_{\sn}] \nonumber\\*
&&\hphantom{\dot{\rho_{\sn}}} + 
\sum_n \frac{\kappa_n (\bar{n}_n+1)}{2} \mathcal{D}[b_n] \rho_{\sn} + 
\sum_n  \frac{\kappa_n \bar{n}_n}{2} \mathcal{D}[b_n^\dagger] \rho_{\sn}, 
\nonumber\\* 
\label{eqn:sigma_phonon_ME}
\end{eqnarray}

\noindent where

\begin{eqnarray}
&&H_{\sn} = -\frac{1}{2}\sum_l \delta_R \sigma_l^z + 
\sum_n \omega_n^\prime b_n^\dagger b_n \nonumber\\*
&&\hphantom{H_{\sn} =} 
+ \sum_{l} \sum_{n} \left( \mathcal{F}_{ln} \sigma_l^+ b_n + \text{h.c.} 
\right).
\end{eqnarray}

Here, $\rho_{\sn}$ is the density matrix describing the $\sigma$-spins and 
the normal modes.

It is convenient to first transform to an interaction picture with 
$H_0 = -\delta_R \left( \frac{1}{2}\sum_l \sigma_l^z + 
\sum_n b_n^\dagger b_n \right)$. The Hamiltonian appearing in Eq. 
(\ref{eqn:sigma_phonon_ME}) in this interaction picture is

\begin{equation}
H_I = \sum_n \td{\delta}_n b_n^\dagger b_n 
+ \sum_{l} \sum_{n} \left( \mathcal{F}_{ln} \sigma_l^+ b_n + \text{h.c.} 
\right),
\end{equation}

\noindent where $\td{\delta}_n = \omega_n^\prime + \delta_R$ is the effective 
detuning of the normal mode $n$. We assume that the Raman laser beams are 
tuned very close to the highest frequency mode, which we take to be the 
center-of-mass (COM) mode, so that $\delta_R \approx -\omega_{\COM}$. 
As a result, $|\td{\delta}_n|$ is very small for the COM mode and increases 
with decreasing mode frequency.

The Liouvillian in Eq.~(\ref{eqn:sigma_phonon_ME}) can be split into a term 
$\lv_R$ acting on the reservoir of normal modes and a term $\lv_{SR}$ that 
couples the system of $\sigma$-spins with this reservoir:

\begin{eqnarray}
&&\dot{\rho}_{\sn} = \lv_R \rho_{\sn} + \lv_{SR} \rho_{\sn}, \quad 
\text{where} \nonumber\\*
&&\lv_R \rho_{\sn} = -i \left[ \sum_n \td{\delta}_n b_n^\dagger b_n, 
\rho_{\sn} \right] \nonumber\\*
&&\hphantom{\lv_R \rho} 
+ \sum_n \frac{\kappa_n (\bar{n}_n+1)}{2} \mathcal{D}[b_n] \rho_{\sn} + 
\sum_n  \frac{\kappa_n \bar{n}_n}{2} \mathcal{D}[b_n^\dagger] \rho_{\sn}, 
\nonumber\\*
&&\lv_{SR} \rho_{\sn} = 
-i \left[ \sum_{l} \sum_{n} \left( \mathcal{F}_{ln} \sigma_l^+ b_n 
+ \text{h.c.}\right), \rho_{\sn} \right].
\label{eqn:sigma_phonon_BM_start}
\end{eqnarray}

The spin-spin interactions are mediated predominantly by the nearly-resonant 
COM mode. If the damping rate $\kappa_{\COM}$ of the COM mode is large 
compared to the collectively-enhanced spontaneous emission rate 
$N_\sigma \Gamma_{\COM} (1 + \nbar_{\COM})$, 
with $\Gamma_{\COM} = \mathcal{F}^2_{\COM}/\kappa_{\COM}$, we can obtain 
an effective master equation for the spin dynamics using second-order 
perturbation theory and a Markov approximation. The details of this 
procedure, and an explanation for the validity condition mentioned above 
are presented in Appendix~\ref{app:eff_spin_spin_model}. The off-resonant 
modes are detuned by $\td{\delta}_n > \kappa_{\COM}$, ensuring the 
Markov approximation can be used for the off-resonant modes as well 
while studying the system on timescales $t \gg \kappa_{\COM}$.

The damping of the normal modes leads to dissipation of energy from the 
system. To maintain steady-state, energy is replenished by continuous 
incoherent repumping of the $\sigma$-spins at a rate $w$. This can be 
achieved by driving the $\ket{1}$ state to an excited state $\ket{a}$, which 
then rapidly decays to $\ket{3}$. The effective master equation for the 
density matrix $\mu_{\sigsub}$ of the $\sigma$-spins, interacting with a 
damped set of normal modes and being incoherently repumped is given by

\begin{eqnarray}
&&\dot{\mu}_{\sigsub} = -i[H_{\sigsub}^{\text{eff}},\mu_{\sigsub}] 
\nonumber\\*
&&\hphantom{\dot{\mu_{\sigsub}} =} 
+ \sum_{l,m} \Gamma_{lm}^- (2 \sigma_m^- \mu_{\sigsub} \sigma_l^+ 
- \sigma_l^+ \sigma_m^- \mu_{\sigsub} - \mu_{\sigsub} \sigma_l^+ \sigma_m^-) 
\nonumber\\*
&&\hphantom{\dot{\mu}_{\sigsub} =} 
+ \sum_{l,m} \Gamma_{lm}^+ (2 \sigma_l^+ \mu_{\sigsub} \sigma_m^- 
- \sigma_m^- \sigma_l^+ \mu_{\sigsub} - \mu_{\sigsub} \sigma_m^- \sigma_l^+) 
\nonumber\\* 
&&\hphantom{\dot{\mu}_{\sigsub} =} 
+ \frac{\Gamma_{31}}{2} \sum_l \mathcal{D}[\sigma_l^-] \mu_{\sigsub} 
+ \left( \frac{w +\Gamma_{13}}{2}\right) \sum_l \mathcal{D}[\sigma_l^+] 
\mu_{\sigsub} \nonumber\\* 
&&\hphantom{\dot{\mu}_{\sigsub} =} 
+ \frac{\Gamma_d}{8} \sum_l \mathcal{D}[\sigma_l^z] \mu_{\sigsub}, 
\label{eqn:sigma_sigma_eff_ME}
\end{eqnarray}

where

\begin{equation}
H_{\sigsub}^{\text{eff}} = \frac{1}{2} \sum_l B_l \sigma_l^z 
+ \sum_{\substack{l,m\\l \neq m}} J_{lm} \sigma_l^+ \sigma_m^-. 
\end{equation}

The expressions for the coefficients introduced in 
Eq.~(\ref{eqn:sigma_sigma_eff_ME}) are as follows \cite{bermudezPRL2013}:

\begin{eqnarray}
&&B_l = 
- \sum_n \frac{|\mathcal{F}_{ln}|^2}{\frac{\kappa_n^2}{4} + \td{\delta}_n^2} 
\td{\delta}_n (1+2\nbar_n), \nonumber\\*
&&J_{lm} = - \sum_n \frac{\mathcal{F}_{ln}\mathcal{F}_{mn}^*}
{\frac{\kappa_n^2}{4} + \td{\delta}_n^2} \td{\delta}_n, \nonumber\\*
&&\Gamma_{lm}^- = \sum_n \frac{\mathcal{F}_{ln}\mathcal{F}_{mn}^*}
{\frac{\kappa_n^2}{4} + \td{\delta}_n^2} \frac{\kappa_n}{2} (1+\nbar_n), 
\nonumber\\*
&&\Gamma_{lm}^+ = \sum_n \frac{\mathcal{F}_{ln}\mathcal{F}_{mn}^*}
{\frac{\kappa_n^2}{4} + \td{\delta}_n^2} \frac{\kappa_n}{2} \nbar_n.
\label{eqn:spin_spin_coeffs}
\end{eqnarray}

It is useful to gain some physical insight into the terms present in the 
master equation (\ref{eqn:sigma_sigma_eff_ME}). The terms 
$\sum_{l,m} \Gamma_{l,m}^- (2 \sigma_m^- \mu_{\sigsub} \sigma_l^+ - \ldots)$ 
and 
$\sum_{l,m} \Gamma_{l,m}^+ (2 \sigma_l^+ \mu_{\sigsub} \sigma_m^- - \ldots)$ 
describe \emph{collective} emission and \emph{collective} absorption of the 
spins respectively. The emission is stronger than the absorption when the 
modes are continuously cooled; this is reflected in the expressions for 
$\Gamma_{l,m}^-, \Gamma_{l,m}^+$ in Eq.~(\ref{eqn:spin_spin_coeffs}). 
The terms of the form $\mathcal{D}[\sigma_l^-] \mu_{\sigsub}$, 
$\mathcal{D}[\sigma_l^+] \mu_{\sigsub}$ and 
$\mathcal{D}[\sigma_l^z] \mu_{\sigsub}$ decribe 
spontaneous emission, incoherent repumping and dephasing respectively. 
The Hamiltonian terms arise because of couplings mediated by the 
off-resonant normal modes; note that the expressions for $B_l$ and $J_{l,m}$ 
vanish when the detunings of all the modes are zero.  The Hamiltonian terms 
comprise an effective magnetic field $B_l$ for each spin, as well as 
pair-wise spin-spin interactions which swap the excitation between the spins.   

Eq.~(\ref{eqn:sigma_sigma_eff_ME}) reveals that the ion trap model has the key 
ingredients required to capture steady-state superradiance: collective 
emission and incoherent repumping. In addition, the ion trap model also 
replicates the spontaneous emission and dephasing processes that may arise 
with neutral atoms in a cavity.

There are two important differences between the steady-state 
superradiance models in an ion-trap and in a cavity. Firstly, the ion 
trap model also has a collective absorption process, which is present 
because of the non-zero temperature set by the Doppler cooling. Further, 
there are Hamiltonian interactions that are mediated by the off-resonant 
normal modes. This feature is absent in the cavity model where it is usually 
a good approximation to consider just a single optical mode. In spite of this, 
the qualitative features of the dynamics in the ion 
trap model are the same as in the cavity model, as we demonstrate in the 
next section.

\section{\label{sec:a_model_system}A model system}

\subsection{Trap, ions and laser configurations}

We first set the stage by considering a concrete example of an ion trap 
system. We consider two species of ions, $^{24}\text{Mg}^+$ and 
$^{25}\text{Mg}^+$, loaded in a Penning trap. The Penning trap allows for 
controlling large numbers of ions, and also gives a well
separated center-of-mass (COM) mode \cite{sawyerPRL2012}
(tens of kilohertz higher than subsequent mode)
that makes it possible to mediate 
superradiant interactions predominantly through a single bosonic mode, 
as in the cavity case. 

Penning traps employ static 
electric fields and a strong uniform magnetic field $\vec{B} = B \hat{z}$ to 
confine ions \cite{biercukQIC2009}. 
The static electric fields are generated by applying 
potentials to electrodes with a common symmetry axis that is aligned with 
the magnetic field ($\hat{z}$) axis. The electric fields provide harmonic 
confinement along the $z$-axis characterized by a transverse frequency 
$\omega_{\COM}$ (this is the frequency of the center-of-mass (COM) mode, 
which is also the highest frequency mode). The combination of the electric 
and magnetic fields leads to $\vec{E} \times \vec{B}$ drift of the ions 
around the $z$-axis. This rotation provides the necessary radial 
confinement. Additional segmented electrodes can be used to apply a rotating 
potential (`rotating wall'), and the rotation of the ions can be 
phase-locked to this `rotating wall' potential, lending stability to the 
system. For sufficiently weak radial confinement, the ions form a 2D planar 
crystal with a triangular lattice, as indicated in 
Fig.~\ref{fig:ion_trap_mapping}. For our model parameters, we set the 
transverse frequency $\omega_{\COM}/2\pi = 2 \; \text{MHz}$, and the lattice 
spacing between adjacent ions to be $a = 10 \; \text{$\mu$m}$. This is 
possible with a transverse magnetic field of $B \approx 5 \; \text{T}$.

\begin{figure}[!htb]
\centering
\includegraphics[width=\linewidth]
{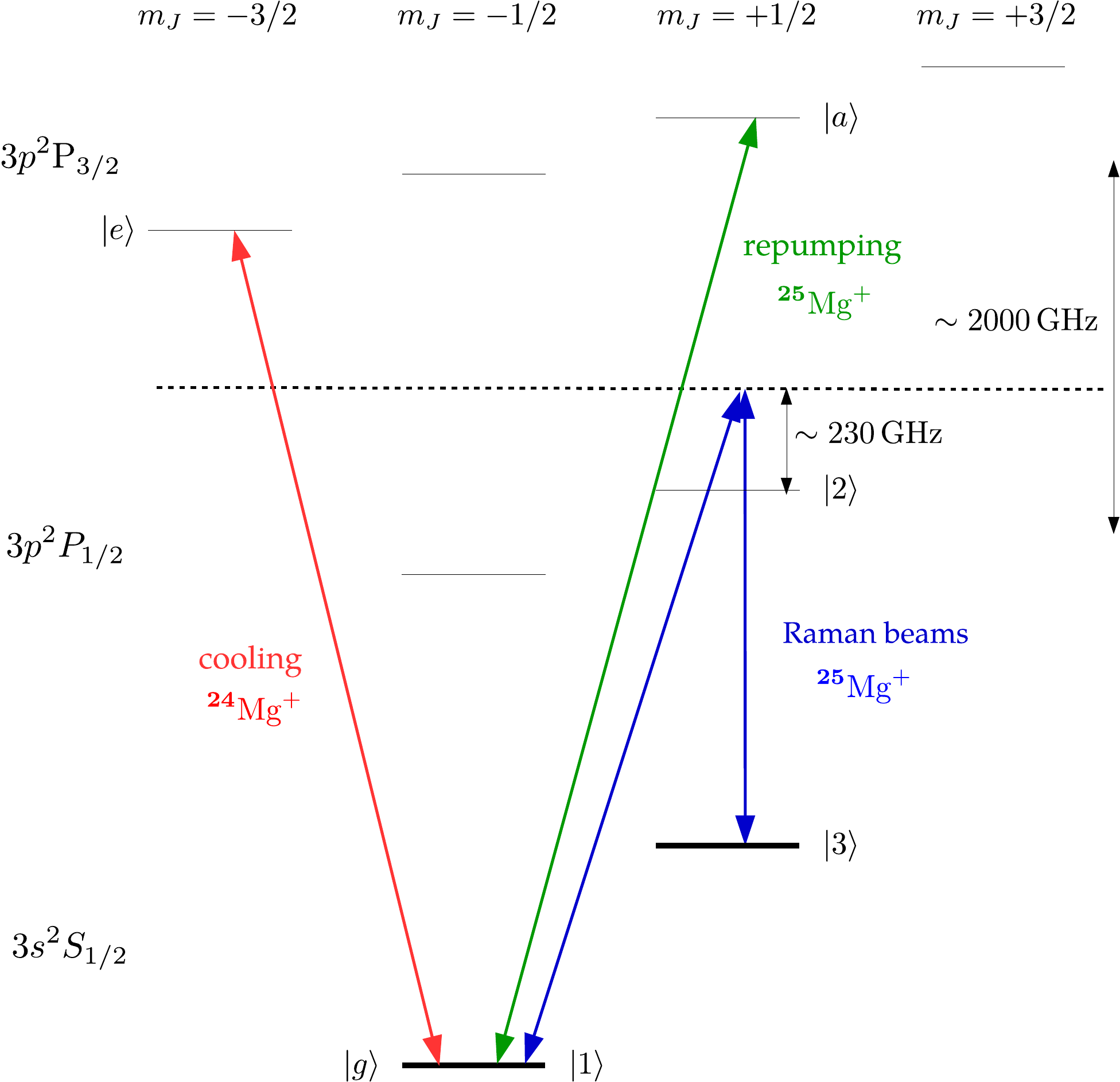}
\caption{(color online) Level structure of $^{24}\text{Mg}^+$ and $^{25}\text{Mg}^+$ ions 
in high field \cite{itanoPRA1981,[See ] [ for precision spectroscopy 
measurements of $\text{Mg}^+$ in zero magnetic field.] batteigerPRA2009} . 
The hyperfine shifts between the two species are not shown 
here. The laser configurations to be used are also indicated. The repump 
laser drives the $3s^2S_{1/2} (m_J = -1/2) \leftrightarrow 
3p^2P_{3/2} (m_J = +1/2)$ transition in $^{25}\text{Mg}^+$, and the 
upper state rapidly decays to 
$3s^2S_{1/2} (m_J = -1/2)$ and $3s^2S_{1/2} (m_J = +1/2)$ with branching 
ratios of $1/3$ and $2/3$ respectively.}
\label{fig:Mg_ion_laser_scheme}
\end{figure}

The centrifugal force brought about by the rotation, causing the heavier ions 
to move outwards, enables separating the two species for different 
functions of the system. The $^{24}\text{Mg}^+$ ions, to be used for 
Doppler cooling ($\tau$ ions), are located in the center, while the 
$^{25}\text{Mg}^+$ ions, to be used as effective spin-$1/2$ systems 
($\sigma$ ions), form hexagonal rings around the inner core of cooling ions. 
In the high magnetic field regime of the Penning trap, the nuclear spin 
$I$ essentially decouples from the 
electronic spin $J$, and $\{J, m_J\}$ are good quantum numbers to describe 
the state of the ions. The level structure of these ions, as well as the 
laser configurations to be used are shown in 
Fig.~\ref{fig:Mg_ion_laser_scheme}.

\subsubsection{$^{24}\text{Mg}^+$ ($\tau$ ions)}

A standing-wave cooling laser 
($\sigma^-$ polarization) is used to drive the 
$3s^2S_{1/2} (m_J = -1/2) \leftrightarrow 3p^2P_{3/2} (m_J = -3/2)$ 
transition ($\ket{g} \leftrightarrow \ket{e}$) which has a separation of 
$\sim 280.3 \; \text{nm}$. The upper level decays at a rate $\Gamma_\tau/2\pi 
\approx 41.4 \; \text{MHz}$ back to the lower level, thereby providing a 
cycling transition for Doppler cooling. The cooling laser has a detuning 
$\Delta_\tau = -\Gamma_\tau/2$ to obtain fast cooling rates. Using a Rabi 
frequency of $\Omega_\tau/2\pi = 10 \; \text{MHz}$ gives a 
cooling rate of $\kappa_{\COM}/2\pi \sim 5 - 6 \; \text{kHz}$ and a mean 
occupation $\bar{n}_{\COM} \approx 4.7$ for the COM mode. 

\subsubsection{$^{25}\text{Mg}^+$ ($\sigma$ ions)}

Two Raman beams (Rabi frequencies 
$|g_1|/2\pi = |g_2|/2\pi \approx 44.7 \; \text{MHz}$), with $\pi$ and 
$\sigma^+$ polarizations respectively couple the 
$3s^2S_{1/2} (m_J = +1/2) \; (\ket{3})$ and the 
$3s^2S_{1/2} (m_J = -1/2) \; (\ket{1})$ levels to the 
$3s^2P_{1/2} (m_J = +1/2) \; (\ket{2})$ level in a far detuned regime 
($\Delta \approx 230 \; \text{GHz}$). Their 
difference detunings are chosen such that $\delta_R 
\approx -\omega_{\COM}^\prime$, where $\omega_{\COM}^\prime$ is the frequency 
of the COM mode, slightly shifted in the presence of the Doppler 
cooling. The Raman beams are oriented such that the Lamb-Dicke parameter for 
the COM mode is $\eta_{\COM}^\sigma \approx 0.1$. A repump laser ($\sigma^+$ 
polarization) drives the $3s^2S_{1/2} (m_J = -1/2) \leftrightarrow 
3p^2P_{3/2} (m_J = +1/2)$ transition ($\ket{1} \leftrightarrow \ket{a}$), 
and the upper level rapidly decays to 
$\ket{1}$ and $\ket{3}$ with a \emph{relative} branching 
ratio $\chi$ of $0.5$. Here, $\chi$ is the ratio of the decay rate back to 
the level $\ket{1}$ and the decay rate to the level $\ket{3}$.  
To illustrate the important physics, the branching back to 
the initial state will be ignored initially; however, we will discuss its 
effects subequently.

We note here that the Raman beams resonantly tuned to interact with the 
$^{25}\text{Mg}^+$ ions will not resonantly interact with the 
$^{24}\text{Mg}^+$ ions; the $^{25}\text{Mg}^+$ ions have a non-zero 
nuclear spin $\vec{I}$ leading to a hyperfine perturbation $A m_I m_J$ that 
changes the level spacing of the effective two-level system by a 
few gigahertz \cite{itanoPRA1981}.

\subsection{\label{subsec:results}Results from numerical simulation}

In a cavity system, steady-state superradiance can be observed 
experimentally by measuring the intensity (photons) and phase 
properties of the output light from the cavity \cite{bohnetNat2012}. 
The corresponding observables in an ion trap are the intensity  
(phonons) and oscillation phase of the COM mode. While in principle 
measurable \cite{biercukNat2010,hempelNat2013}, factors like the 
background phonons from Doppler cooling have to be carefully considered 
before embarking on such measurements. Standard techniques in ion traps 
offer convenient ways to directly study the spin degrees of freedom. 
Steady-state superradiance is characterized by the development of 
non-zero steady-state spin-spin correlations, leading to the formation 
of a giant collective spin which 
behaves very differently compared to uncorrelated spins. It is this aspect 
of superradiance that we study numerically and propose techniques for 
probing via experiments. 

We define an ensemble-averaged (EA) rate 
$\Gamma_c = (2/N_\sigma^2) \sum_{l,m} ( \Gamma_{lm}^- - \Gamma_{lm}^+ )$, 
which plays an analogous role to the net single-atom emission rate 
into the cavity mode in the superradiant laser \cite{meiserPRA2010}. 
The strength of the Raman beams have been chosen such that 
the nearly resonant COM mode is strongly damped compared to the 
collectively-enhanced spontaneous emission rate, i.e.  
$\kappa_{\COM} \gg N_\sigma \Gamma_c (1+ \nbar_{\COM})$.
Steady-state superradiance is expected in a regime 
where the repump strength $w \lesssim N_\sigma \Gamma_c$ 
\cite{meiserPRA2010}. We are interested in the collective 
behavior of a large number of ions; however, 
the exact solution is near impossible to compute since the density matrix 
lives in a $4^{N_\sigma}$ dimensional Hilbert space, limiting computation of 
exact solutions of the master equation to $N_\sigma \lesssim 10$. 
We use an approximate technique using $c$-number Langevin equations 
to analyze this problem. This 
involves writing the quantum Langevin equations for the spin operators 
$\sigma_l^\pm, \sigma_l^z$ using the master equation 
(\ref{eqn:sigma_sigma_eff_ME}), obtaining the noise correlations using the 
Einstein relations \cite{meystre1998elements}, and finally 
making a correspondence between quantum 
operators and classical $c$-numbers in order to obtain $c$-number Langevin 
equations. This is elaborated in 
Appendix~\ref{app:c_number_langevin}. Table~\ref{tab:model_params} 
gives the important parameters for numerical simulation of a system 
comprising $N_\sigma = 124$ and $N_\tau = 93$ ions.

\begin{table}
\caption{\label{tab:model_params}(color online) Summary of important parameters for a 
numerical simulation, for a system consisting of $N_\sigma = 124$ and 
$N_\tau = 93$ ions (giving a total of $N = 217$ ions). The table also 
shows the ion positions used for numerical simulation. }
\begin{ruledtabular}
\begin{tabular}{>{\quad}p{0.55\linewidth}p{0.35\linewidth}}
\hspace{-1 em} 1. Trap \\
\hspace{-1 em} Input parameters \\
a. System size $(N_\sigma, N)$ & $(124,217)$ \\
b. Lattice spacing & $10 \; \mu\text{m}$\\ 
c. COM mode frequency & $2 \; \text{MHz}$\\
d. Ion positions (orange: $\sigma$, blue: $\tau$) & \raisebox{-\height}
{\includegraphics[scale=0.2]{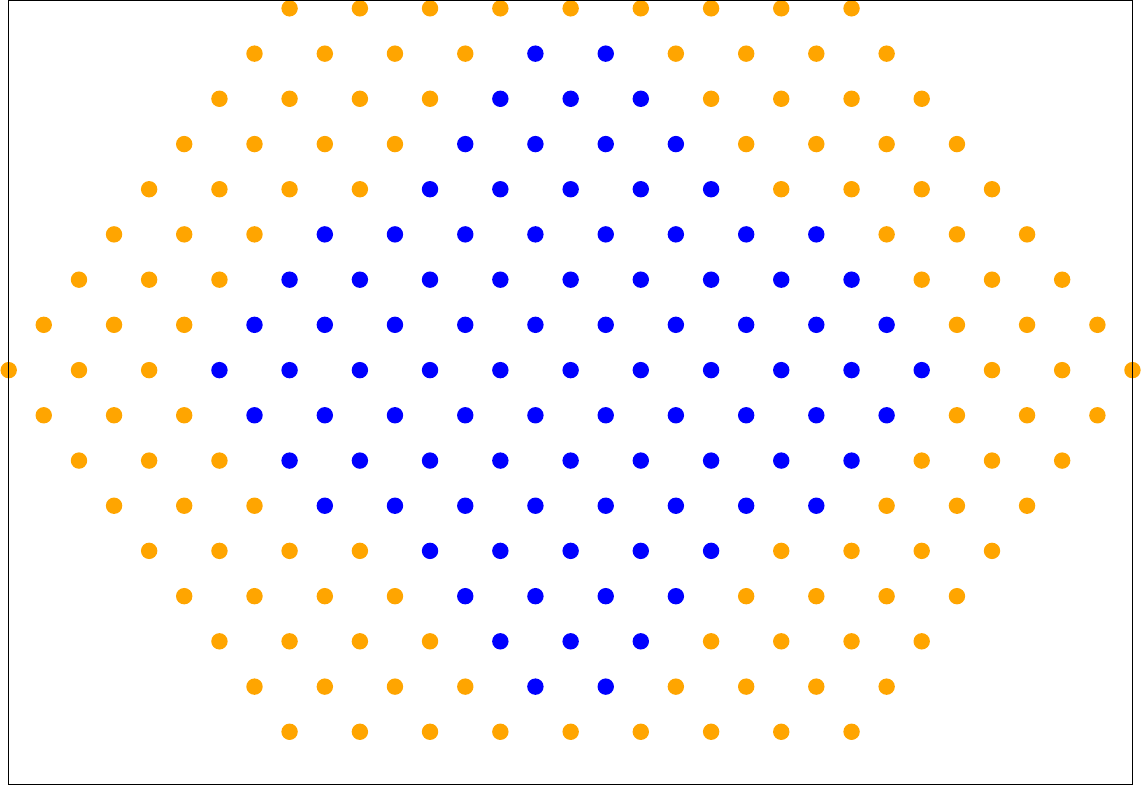}}\\
\hspace{-1 em} Derived parameters \\
a. Other normal modes & diagonalize potential energy matrix for above geometry.
\\ \hline
\hspace{-1 em} 2. $\tau$ ions \\
\hspace{-1 em} Input parameters \\
a. Upper level decay rate $\Gamma_\tau$ & $41.4 \; (2\pi \times \text{MHz})$ \\
b. Transition wavelength & $280.3 \; (\text{nm})$ \\
c. Cooling laser detuning $\Delta_\tau$ & $-\Gamma_\tau/2$ \\
d. Cooling laser Rabi freq. $\Omega_\tau$ & $10 \; (2\pi \times \text{MHz})$\\  
\hspace{-1 em} Derived parameters \\
a. Cooling rate $\kappa_{\COM}$ & $5.1 (2\pi \times \text{kHz})$\\
b. Mean occupation $\nbar_{\COM}$ & $4.7$\\ \hline
\hspace{-1 em} 3. $\sigma$ ions \\
\hspace{-1 em} Input parameters \\
a. Raman beams $g_1 = g_2$ & $44.7 \; (2\pi \times \text{MHz})$\\
b. Average detuning $\Delta$ & $230 \; (2\pi \times \text{GHz})$ \\
c. Difference detuning $\delta_R$ & $-\omega_{\COM}^\prime$ \\
d. Lamb-Dicke parameter $\eta_{\COM}$ 
(sets difference wavevector $|k_\sigma|$) & $0.1$ \\
e. Scattering from $\ket{2}$: $\Gamma_1$, $\Gamma_2$ & $27.27, 13.63 \; 
(2\pi \times \text{MHz})$\\ 
f. Repump $w$ & variable; $0.05-1.0 \; N_\sigma \Gamma_c$ \\
\hspace{-1 em} Derived parameters \\
a. Coupling constants $B_l$, $J_{lm}$, $\Gamma_{lm}^\pm$ & calculate from 
Eq.~(\ref{eqn:spin_spin_coeffs}) \\
b. Net collective emission rate $\Gamma_c$ & $0.84 \; (2\pi \times \text{Hz})$ 
\\
c. Spontaneous Raman $\Gamma_{13}$, $\Gamma_{31}$, $\Gamma_d$ & 
$0.12, 0.24, 0.36 \; (2\pi \times \text{Hz})$\\
\end{tabular}
\end{ruledtabular}
\end{table}

\subsubsection{Steady-state inversion and spin-spin correlation}

\begin{figure}[!htb]
\centering
\subfigimg[width=\linewidth]{\large(a)}{40pt}
{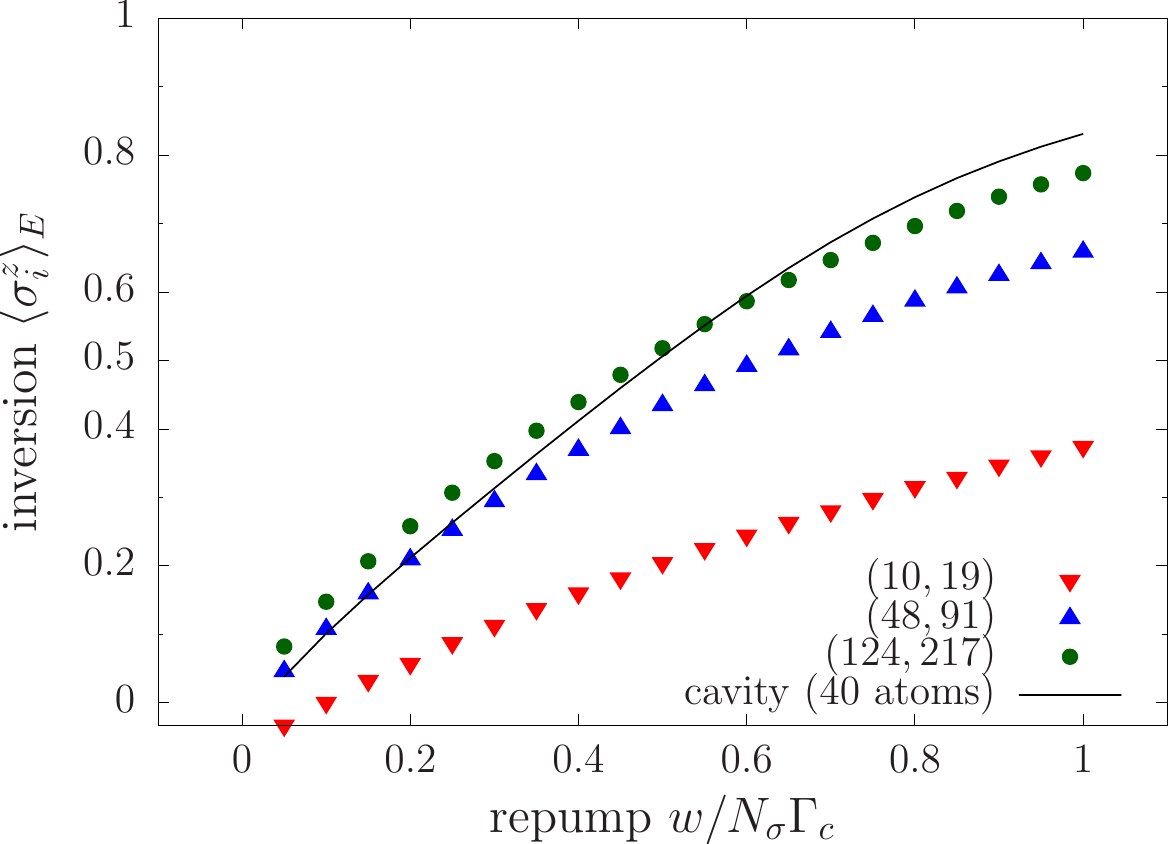}\\[10pt]
\subfigimg[width=\linewidth]{\large(b)}{43pt}
{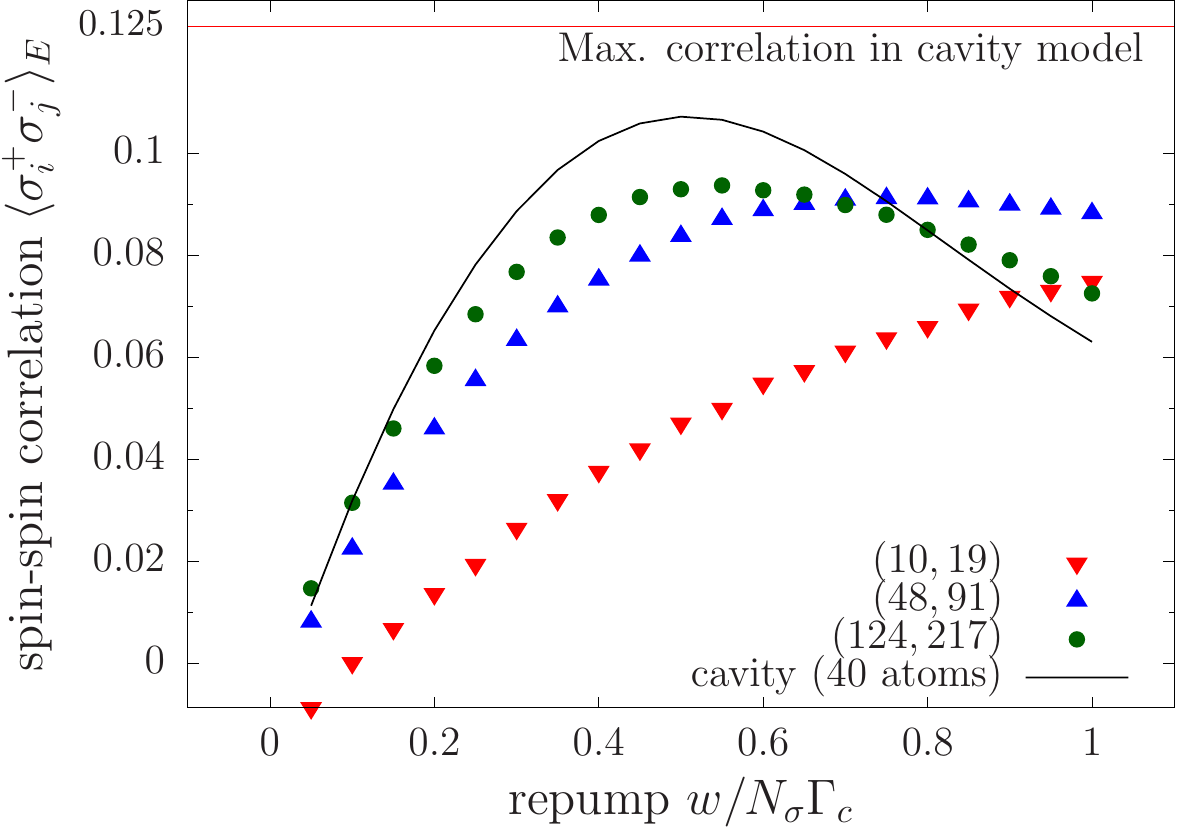}
\caption{(color online) Steady-state (a) Inversion and (b) spin-spin correlation as a 
function of repump strength for three different system sizes (SS) 
($N_\sigma$, $N$). The corresponding values for a minimal cavity 
model with $N_\sigma = 40$ atoms are also plotted. As the system size 
increases, the inversion and correlation in the ion trap case become similar 
to the cavity case. }
\label{fig:inversion_correlation_repump_different_N}
\end{figure}

The system size (SS) can be specified using the notation ($N_\sigma$, $N$), 
where $N = N_\sigma + N_\tau$. An increase in $N_\sigma$ is accompanied by 
an increase in $N_\tau$, because for the same laser power, more coolant ions 
are required to provide fast cooling rates when larger number of ions are 
present. We will use the notation $\langle \ldots \rangle_E$ to denote 
expectation values that are averaged over the entire ensemble of spins. 
Figure~\ref{fig:inversion_correlation_repump_different_N} shows the 
steady-state EA inversion and spin-spin correlation 
($\langle \sigma_i^+ \sigma_j^- \rangle_E$)\footnote{ In steady-state, 
$\langle \sigma_i^\pm\rangle = 0$ for all spins $i$. 
Therefore, the correlation $\langle \sigma_i^+ \sigma_j^- \rangle - 
\langle \sigma_i^+ \rangle \langle \sigma_j^- \rangle$ for every pair 
$i,j$ of spins is simply $\langle \sigma_i^+ \sigma_j^- \rangle$.}
for three different system sizes: 
(i) (10, 19), (ii) (48, 91), and (iii) (124, 217). In a minimal cavity model 
that accounts for only collective emission and incoherent repumping 
\cite{meiserPRA2010_2}, the 
steady-state values do not change significantly for $N_\sigma \gtrsim 30$ 
atoms. The inversion and correlation in the cavity case for $N_\sigma = 40$ 
atoms are shown for comparison. As the system size increases, both the 
inversion and correlation for the ion trap system become similar to the 
cavity case \cite{meiserPRA2010_2}: 
for large $N_\sigma$, the inversion grows monotonically with 
$w$, and is approximately $1/2$ at $w = 0.5 N_\sigma \Gamma_c$ (collective 
Bloch vector is halfway between equator and North Pole). The correlation 
increases with $w$, reaches a maximum around $w = 0.5 N_\sigma \Gamma_c$, 
and then decreases with further increase in $w$. The development of 
steady-state pair-wise spin-spin correlations implies the phase-locking of 
spins, and the formation of a giant collective spin, which is a signature 
of steady-state superradiance. It is reasonable to expect that the ion trap 
system gives results similar to the zero temperature minimal cavity model as 
the system size increases; the corrections to the inversion and correlation, 
due primarily to a non-zero temperature set by $\bar{n}_{\COM}$, scale as 
$\bar{n}_{\COM}/N_\sigma$. This can be seen by estimating the steady-state 
values by writing the equations of motion for these expectation values and 
closing the set of equations by performing a cumulant approximation as 
was done in Ref. \cite{meiserPRA2010, xuPRL2015}.  
 
\subsubsection{Experimental access: Ramsey fringes}

\begin{figure}[!htb]
\centering
\subfigimg[width=\linewidth]{\large(a)}{40pt}
{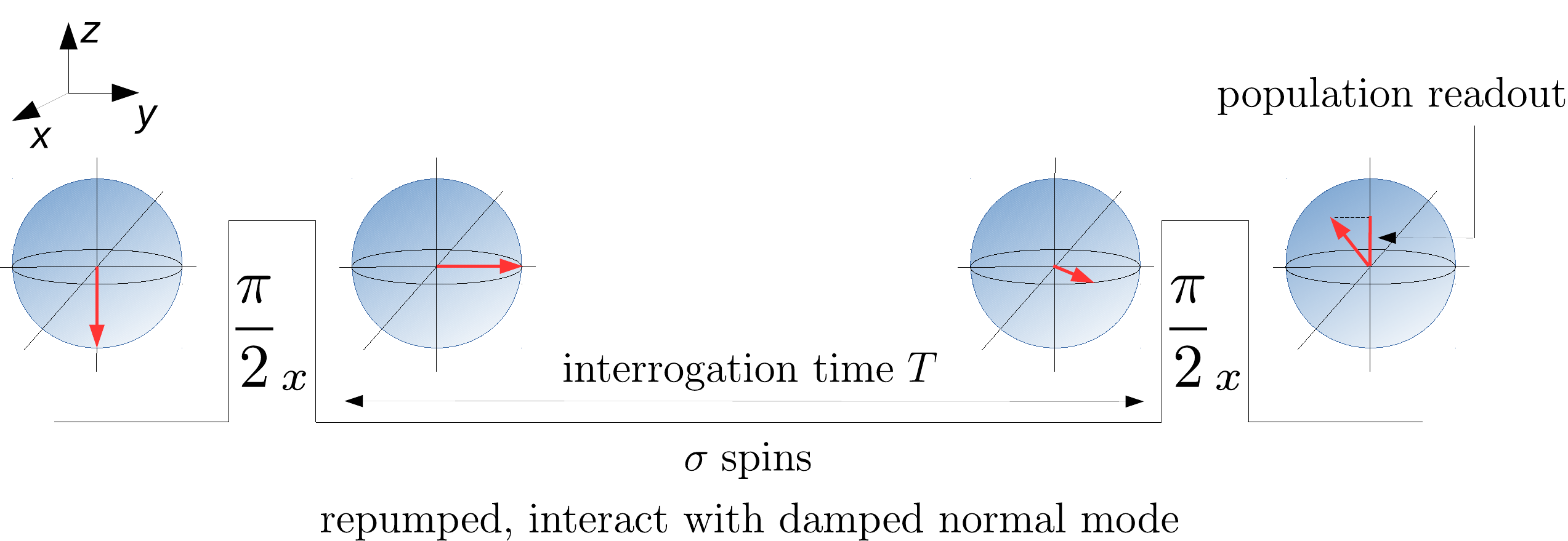}\\[10pt]
\subfigimg[width=\linewidth]{\large(b)}{40pt}
{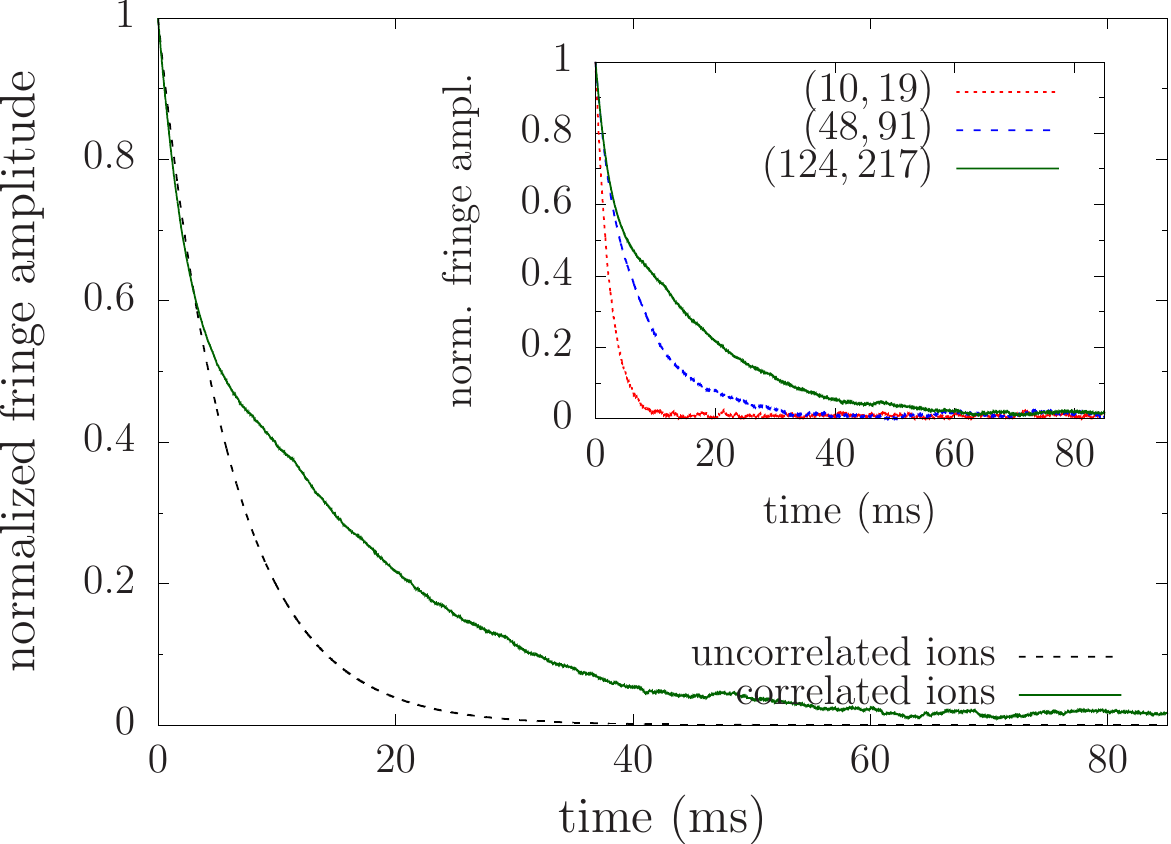}
\caption{(color online) (a) Ramsey pulse sequence to probe the collective spin. During the 
interrogation time, the $\sigma$ ions interact with a heavily damped 
normal mode while being continuously repumped. (b) Decay of the Ramsey 
fringe envelope for uncorrelated ions and a system 
of correlated ions with SS (124, 217) and $w = N_\sigma \Gamma_c/2$. Once 
the $\sigma$-spins have phase-locked, the Ramsey fringe decays at a slower 
rate than when the spins are uncorrelated.  Inset: 
Fringe decay as a function of time for three different system sizes for 
$w = N_\sigma \Gamma_c/2$.}
\label{fig:ramsey_pulse_demo_fringe}
\end{figure}

In order to observe this collective spin experimentally, a Ramsey pulse 
sequence \cite{ramseyPR1950} could be used 
(see Fig.~\ref{fig:ramsey_pulse_demo_fringe}(a)). 
In a traditional Ramsey sequence, the spins 
initially in the ground state (South Pole of Bloch sphere), are brought to 
a uniform superposition of ground and excited states (equator) by applying 
a $\pi/2$-pulse about the $x$-axis. In the frame of the initial laser, the 
spins then precess around the $z$-axis at a rate set by the detuning of the 
laser, for an interrogation period $T$. Finally, a second $\pi/2$-pulse 
rotates the spin about the $x$-axis and the population is read out using 
resonance fluorescence. The result is a sinusoidal variation (`fringe') of 
the population, with the amplitude damped by incoherent processes such as 
spontaneous emission, incoherent repumping and/or dephasing.

Here, after the initial $\pi/2$-pulse, we intend to allow the $\sigma$-spins 
to interact with the damped set of normal modes during the interrogation 
period, while continuously repumping the spins incoherently at a rate $w$ 
\cite{xuPRL2015}. 
This is achieved by continuous Doppler cooling of the $\tau$ ions, and 
applying Raman and repump beams to the $\sigma$-spins, during the 
interrogation period. Finally, the second $\pi/2$-pulse is applied and the 
population is read out. In the presence of only the repump, the amplitude of 
the fringe decays at a rate $w/2$. However, the damped COM mode mediates 
phase-locking of the spins, that leads to a giant collective spin that is 
robust against individual atom incoherent processes. After a fast initial 
transient during which the spins phase-lock, the fringe 
decays at a slower rate; a rate that is set by the phase diffusion of 
this collective spin. The pair-wise spin-spin interactions 
($O(N^2)$ interactions) lead to phase-locking of the spins, while the self 
interactions of the spins ($O(N)$ interactions) are phase-destroying 
processes that result in phase diffusion. 
Figure~\ref{fig:ramsey_pulse_demo_fringe}(b) compares the fringe decays 
for uncorrelated ions and correlated ions. The inset shows the normalized 
Ramsey fringe amplitude for three different system sizes. 

\begin{figure}[!htb]
\centering
\subfigimg[width=\linewidth]{\large(a)}{40pt}
{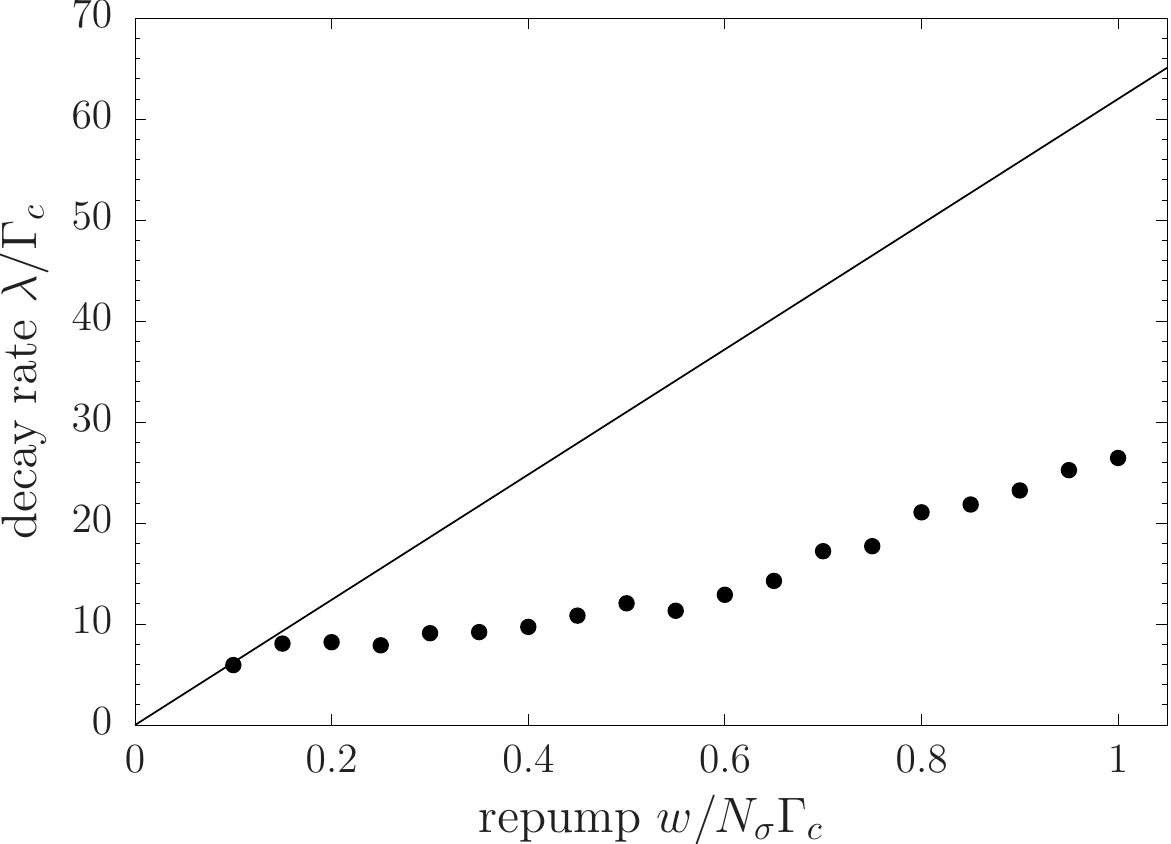}\\[10pt]
\subfigimg[width=\linewidth]{\large(b)}{43pt}
{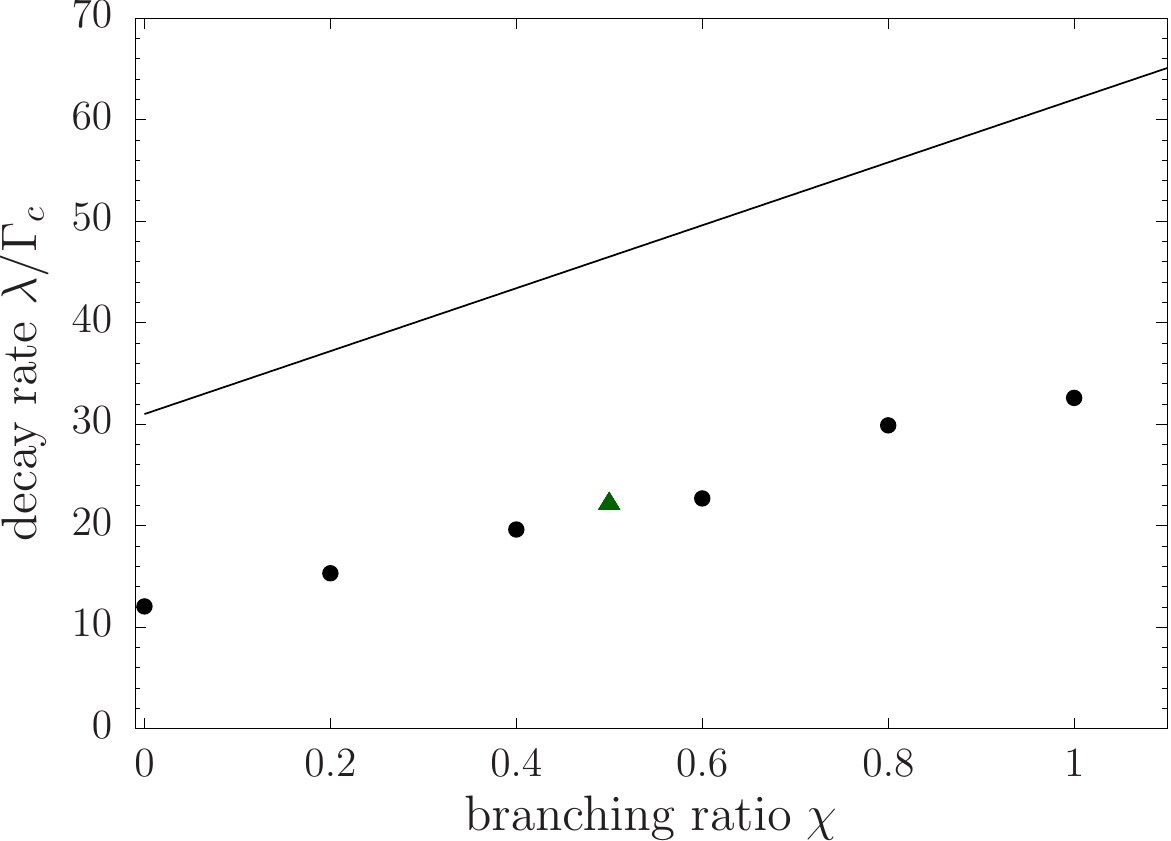}
\caption{(color online) (a) Decay rate of the Ramsey fringe envelope (dots) as a function 
of repump strength for SS (124, 217). The solid line shows the decay rate 
if only repumping is present. (b) Decay rate of the Ramsey fringe envelope 
(dots) as a function of the relative branching ratio for SS (124, 217) and 
$w = N_\sigma \Gamma_c/2$. The solid line shows the decay rate if only 
repumping (with branching) is present. The repumping scheme proposed in
this paper (see Fig.~\ref{fig:Mg_ion_laser_scheme}) 
with the $^{25}\text{Mg}^+$ ions has a relative
branching ratio of $0.5$, and is indicated by a green triangle. }
\label{fig:decay_rate_repump_branching_ratio_N124}
\end{figure}

Figure~\ref{fig:decay_rate_repump_branching_ratio_N124}(a) shows the decay 
rate of the Ramsey fringe envelope as a function of repump strength for 
SS (124, 217). The collective spin clearly decays at a slower rate compared 
to the case when only repumping is present, indicating phase-locking of the 
spins. 

In contrast to simple repumping schemes (Fig.~\ref{fig:sigma_ion_level}), 
the excited state $\ket{a}$ does not rapidly decay to 
$\ket{3}$ alone in realistic repumping schemes. A fraction of the 
population in $\ket{a}$ also decays back to the initial state $\ket{1}$, 
with a relative branching ratio $\chi$ that gives the ratio of population 
transfer to $\ket{1}$ and $\ket{3}$. The effect of this is to introduce 
an additional dephasing $\Gamma_w = \chi w$, where $w$ is the repumping 
strength. This can be accounted for by setting 
$\Gamma_d \rightarrow \Gamma_d + \Gamma_w$ in 
the master equation (\ref{eqn:sigma_sigma_eff_ME}). The decay rate for 
various relative branching ratios is shown in 
Fig.~\ref{fig:decay_rate_repump_branching_ratio_N124}(b) for SS (124, 217) and 
$w = 0.5 N_\sigma \Gamma_c$. The dephasing due to branching scales with the 
rate of synchronization, which is set by the repump strength $w$. Despite 
this, the phase-locking of the spins still ensures that the fringe amplitude 
decays slower compared to the situation when only repumping (with branching) 
is present. 

\begin{figure}[!htb]
\includegraphics[width=\linewidth]
{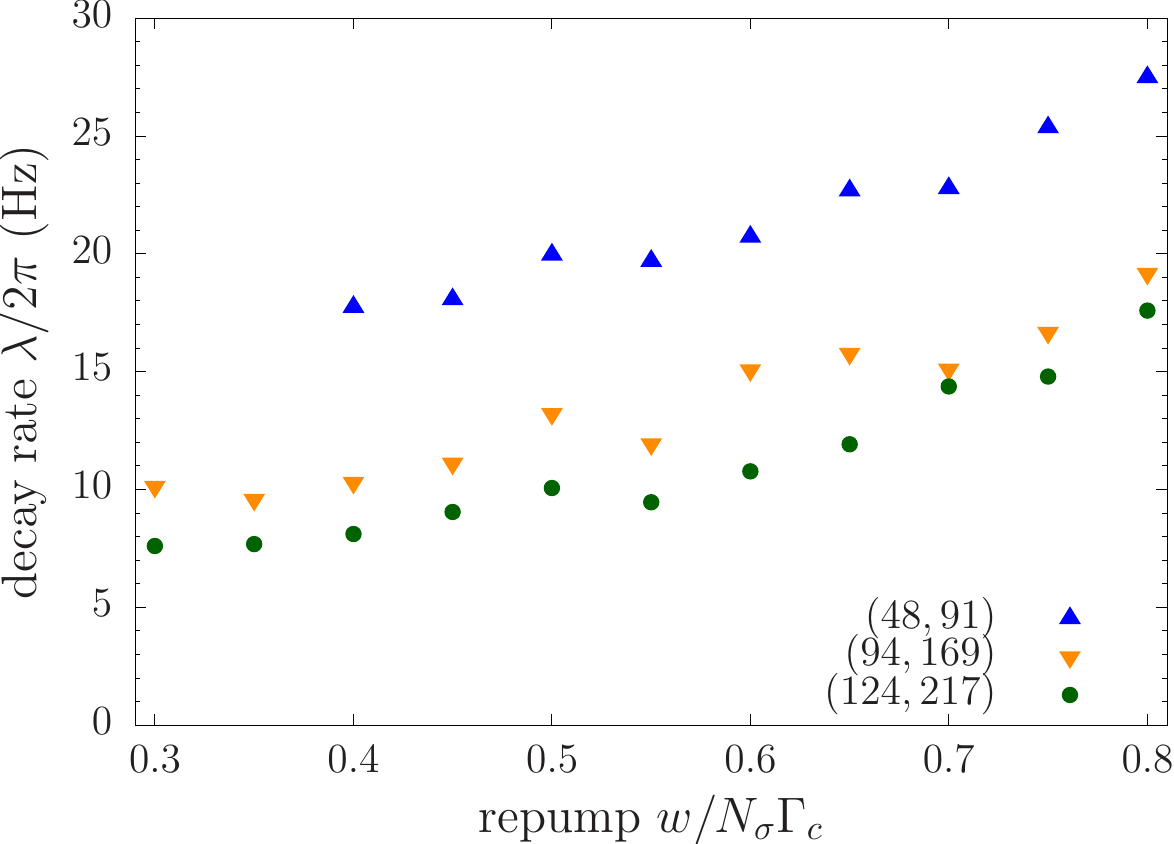}
\caption{(color online) Decay rate of the Ramsey fringe envelope (dots) as a function 
of repump strength for three different system sizes. 
The fringes persist for longer with increasing 
$N$.  }
\label{fig:decay_rate_repump_fringe_time_different_N}
\end{figure}

In the cavity model, the plot of decay rate vs repump strength 
for reasonably large system sizes ($N_\sigma \gtrsim 40$) is approximately 
the same, when the repump strength is in units of $N_\sigma \Gamma_c$. 
However, the constant $\Gamma_c$ is independent of $N_\sigma$ in the cavity 
case. In the ion trap system, the spin-spin coupling is predominantly 
mediated by the single nearly-resonant COM mode, although a total of 
$N$ modes are available. Hence, the coupling of each spin to the COM 
mode scales as $1/\sqrt{N}$, and hence $\Gamma_c$ 
(see Eq.~(\ref{eqn:spin_spin_coeffs})) decreases as $N$ increases. 
As a result, when the decay rate is measured in absolute 
units, say, hertz for example, the decay rate decreases as $N$ increases. 
This is demonstrated in 
Fig.~\ref{fig:decay_rate_repump_fringe_time_different_N}(a) for SS (i) (48, 91), 
(ii) (94, 169), and (iii) (124, 217). 

The variance of the population measurement at the end of the Ramsey sequence 
could give information about the spin-spin correlations present in the 
system. Using $J^x, J^y, J^z$ to denote the components of the collective 
spin, we note that the variance of the total inversion $(\Delta J^z)^2$ after 
the second $\pi/2$-pulse in the Ramsey sequence is just $(\Delta J^y)^2$ 
before that pulse. Before the second $\pi/2$-pulse, the variance 
$(\Delta J^y)^2$ can be expressed as

\begin{eqnarray}
(\Delta J^y)^2 = \frac{N_\sigma}{4} &&+ \frac{N_\sigma(N_\sigma-1)}{2} 
\left( \langle \sigma_i^+ \sigma_j^- \rangle_E - 
\text{Re} \langle \sigma_i^+ \sigma_j^+ \rangle_E \right) \nonumber\\*
&&- N_\sigma^2 \left(\text{Im} \langle \sigma_i^+ \rangle_E\right)^2.
\end{eqnarray} 

\begin{figure}[!htb]
\centering
\includegraphics[width=\linewidth]
{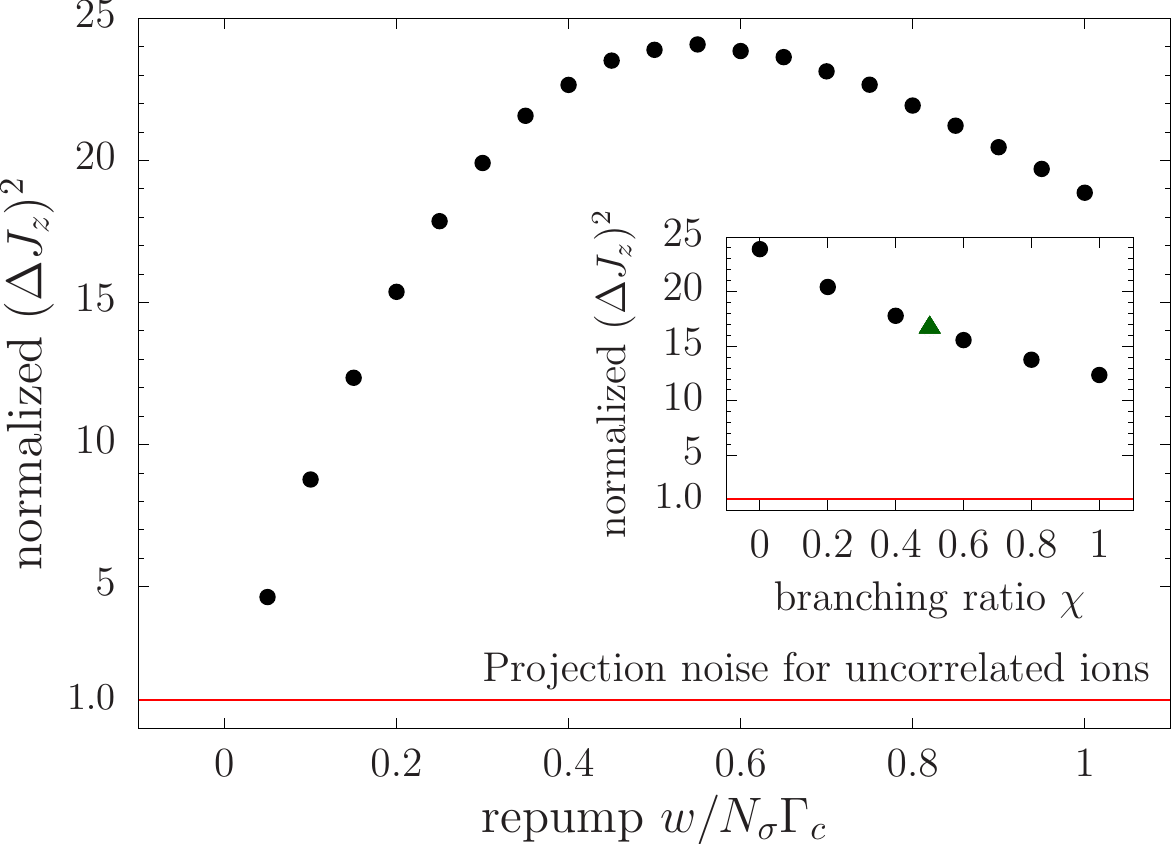}
\caption{(color online) Steady-state variance of inversion as a function of repump 
strength $w$ for SS (124, 217) at the end of the Ramsey pulse sequence, 
normalized to the projection noise for uncorrelated ions ($N_\sigma/4$). 
Inset: Normalized steady-state variance of inversion as a function of 
relative branching ratio $\chi$ for SS (124, 217) and 
$w = N_\sigma \Gamma_c/2$. The repumping scheme proposed 
in this paper (see Fig.~\ref{fig:Mg_ion_laser_scheme}) 
with the $^{25}\text{Mg}^+$ ions 
has a relative branching ratio of 0.5, and is indicated by a green triangle.}  
\label{fig:variance_with_inset_N124}
\end{figure}

The quantities $\text{Re} \langle \sigma_i^+ \sigma_j^+ \rangle_E$ and 
$\text{Im} \langle \sigma_i^+ \rangle_E$ are zero once the fringe envelope 
has decayed to zero. Thus, the steady-state variance $(\Delta J^z)^2$ 
($(\Delta J^y)^2$ before the second $\pi/2$-pulse) scales 
as $N_\sigma^2 \langle \sigma_i^+ \sigma_j^- \rangle_E \; (N_\sigma \gg 1)$ 
giving a measure of the non-zero steady-state spin-spin correlations. 
Experimentally, this corresponds to a situation where the Ramsey fringe 
amplitude has decayed to zero but the variance of the population inversion 
readout is significantly larger ($N_\sigma^2$ scaling) than what we would 
expect for uncorrelated spins, as shown in 
Fig.~\ref{fig:variance_with_inset_N124}. 
Further, the $N_\sigma^2$ scaling shows the all-to-all nature of the 
spin-spin interactions . The variance could be a 
measurable quantity even when the repumping has a non-zero relative 
branching ratio: the inset of Fig.~\ref{fig:variance_with_inset_N124} 
shows the steady-state variance as a function of relative branching ratio 
for SS (124, 217) and $w = N_\sigma \Gamma_c/2$.

\subsubsection{Potential advantage of Sub-Doppler cooling}

\begin{figure}[!htb]
\centering
\subfigimg[width=\linewidth]{\large(a)}{40pt}
{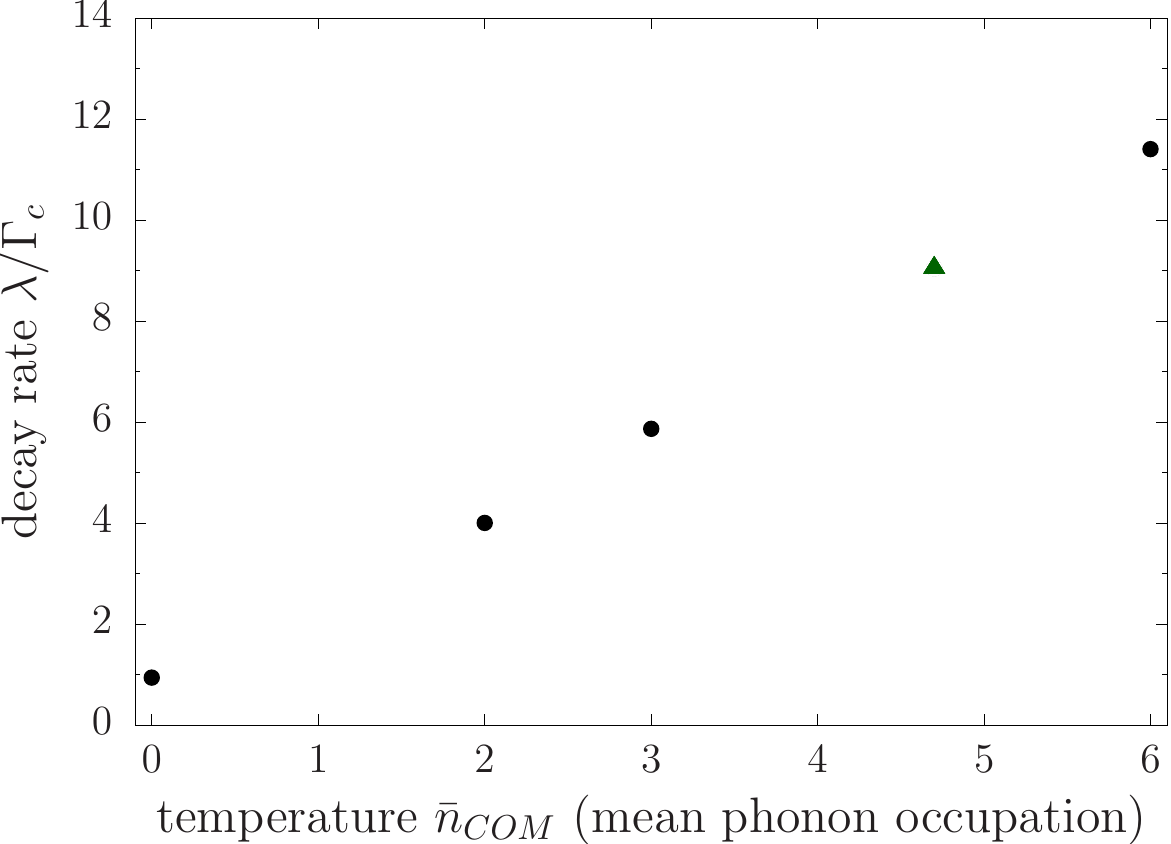}\\[10pt]
\subfigimg[width=\linewidth]{\large(b)}{43pt}
{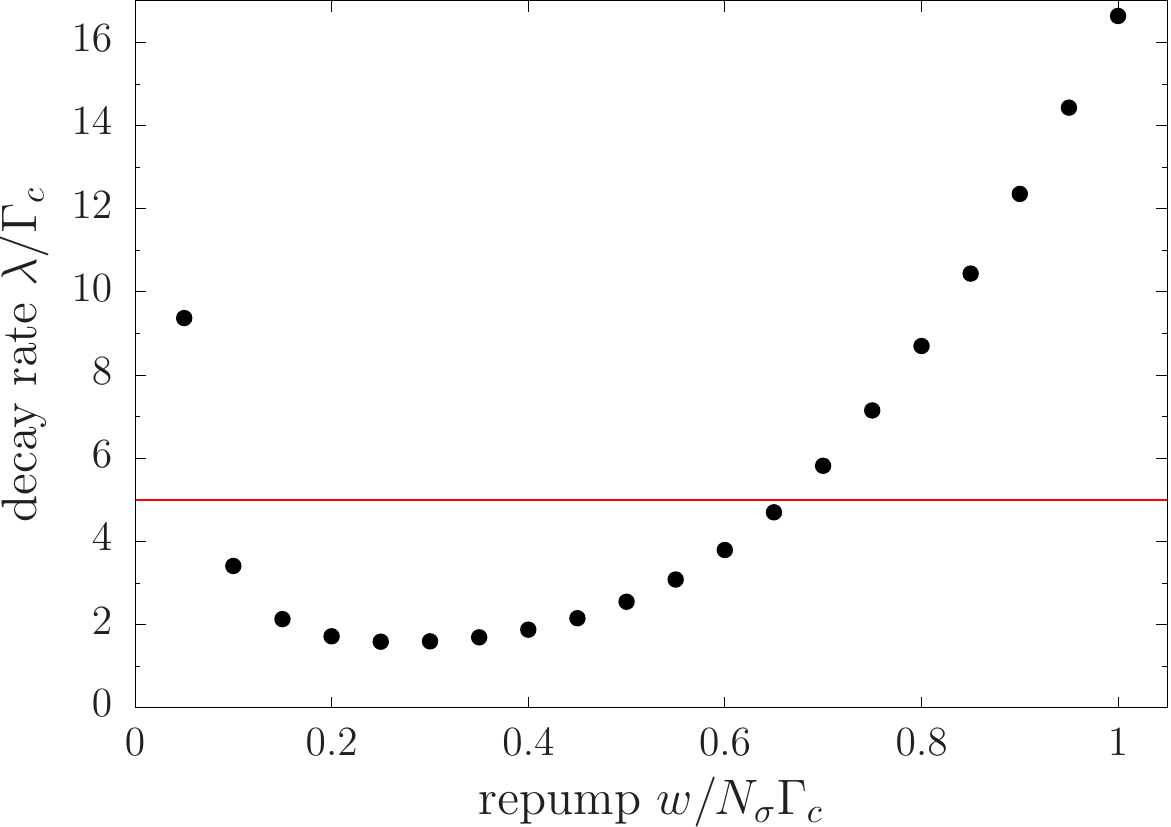}
\caption{(color online) (a) Decay rate of the Ramsey fringe envelope as a function of the 
mean occupation $\bar{n}_{\COM}$ of the center-of-mass (COM) mode for 
$N_\sigma = 124$ ions. The rates shown 
here are calculated using the SU(4) method for a minimal model of a single 
mode (COM) interacting with the $\sigma$ ions (Eq. 
(\ref{eqn:ion_trap_minimal_model})). The Doppler cooling scheme 
proposed is shown by a green triangle. (b) Decay rate of the Ramsey fringe 
envelope as a function of repump strength $w$ for $N_\sigma = 124$ ions. 
The minimal model of Eq.~(\ref{eqn:ion_trap_minimal_model}) is used with 
$\bar{n}_{\COM} = 0$, but with additional spontaneous emission 
and dephasing processes for the individual ions 
($\Gamma_{\text{sp}} =  \Gamma_{\text{d}} = 5 \Gamma_c$). The decay rate 
for uncorrelated ions is shown by the horizontal line 
($\lambda = 5 \Gamma_c$). Synchronization can prolong visibility of Ramsey 
fringes.}
\label{fig:decay_rate_nbar_N124}
\end{figure}

Our current design uses Doppler cooling to provide a heavily damped COM mode 
that can mediate spin-spin interactions. In a minimal model, we can ignore 
the coupling of the spins to all the modes other than the resonant COM mode. 
Further ignoring spontaneous emission and dephasing, this minimal model is 
described by the master equation

\begin{eqnarray}
&&\dot{\mu}_{\sigsub} = 
\frac{\Gamma_c}{2} (\bar{n}_{\COM} + 1) \mathcal{D} [J^-] \mu_{\sigsub} 
+ \frac{\Gamma_c}{2} \bar{n}_{\COM}  \mathcal{D} [J^+] \mu_{\sigsub} 
\nonumber\\*
&&\hphantom{\dot{\mu}_{\sigsub} =} 
+ \frac{w}{2} \sum_l \mathcal{D}[\sigma_l^+] \mu_{\sigsub},
\label{eqn:ion_trap_minimal_model}
\end{eqnarray}

\noindent where $J^\pm = \sum_l \sigma_l^\pm$ are ladder operators for the 
collective spin. It is instructive to study the change in the decay rate as 
$\bar{n}_{\COM}$ is changed. We note that this model is invariant under the 
permutation of spins. We compute exact decay rates of the Ramsey fringes for 
different values of $\bar{n}_{\COM}$ using a numerical method that exploits 
the $SU(4)$ symmetry of spin systems that obey permutation symmetry 
\cite{xuPRA2013}. We 
summarize these results in Fig.~\ref{fig:decay_rate_nbar_N124}(a) for 
SS (124, 217) and $w = N_\sigma \Gamma_c/2$. The decay rate can be as 
low as $\Gamma_c$ if the COM mode is cooled to $\bar{n}_{\COM} \approx 0$. 
With Doppler cooling, our model system has $\bar{n}_{\COM} \approx 4.7$, and 
this gives us a decay rate around $10 \Gamma_c$, an order of magnitude 
higher than what is achievable. Clearly, sub-Doppler cooling techniques 
\cite{[See ] [ for a review of cooling methods in ion traps.] eschnerJOSAB2003, 
monroePRL1995, linPRL2013, lechnerPRA2016}
could be used to observe longer lasting fringes.

Spin synchronization mediated by a sub-Doppler cooled normal mode, and with a 
repumping scheme that has a negligible relative branching ratio, can improve 
metrology with ion traps. With uncorrelated ions that have $1/T_1$ 
(population decay) and $1/T_2$ (dephasing) rates, 
the Ramsey fringe envelope decays at a rate $\Gamma_s =  (T_1^{-1}+T_2^{-1})/2$.
However, with synchronized ions, the Ramsey fringe envelope decays slower 
than in the case of uncorrelated ions in the regime where 
$\Gamma_c \ll \Gamma_s \ll w$ \cite{xuPRL2015}. 
The synchronization effect causes the 
collective spin to be robust against individual ion decoherence processes. 
This is illustrated in Fig.~\ref{fig:decay_rate_nbar_N124}(b) for the minimal
model considered in Eq.~(\ref{eqn:ion_trap_minimal_model}) with 
$\bar{n}_{\COM} = 0$, and with additional spontaneous emission 
($\Gamma_{\text{sp}} = 1/T_1$) and dephasing
($\Gamma_{\text{d}} = 1/T_2$) processes for the individual ions.

\section{\label{sec:conclusion}Conclusion}

We have presented and numerically analyzed a model of steady-state spin 
synchronization in an ion trap, where the synchronization is mediated by a 
heavily damped normal mode of vibration. This is achieved by mapping the 
dynamics of cavity steady-state superradiance onto an ion trap system by 
exploiting the overarching similarity of an optical cavity mode and a normal 
mode of vibration. 

We have considered a model system of two species of ions in a Penning 
trap, although the present scheme can also be implemented with 1D 
or 2D RF traps that can trap a mesoscopic number 
($\gtrsim 20$) of ions. As the system size increases, the steady-state 
spin-spin correlations in the ion trap are similar to that in the cavity 
case, since the effects of a non-zero temperature due to the Doppler cooling 
are negated. 

We have proposed an experimental scheme using a Ramsey pulse sequence that 
can be used to observe features of the collective spin that develops in the 
ion trap. The Ramsey fringes persist longer when the spins are 
synchronized, with a lower decay rate than we expect 
from the incoherent repumping. Further, the variance of the population 
readout at the end of the Ramsey sequence scales as 
$N_\sigma^2 \langle \sigma_i^+ \sigma_j^- \rangle_E$, providing a 
straightforward means for documenting spin-spin correlation and the 
all-to-all nature of the coupling. These signatures of spin synchronization 
persist even when the repumping is imperfect and has a non-zero branching 
ratio back to the initial state. We also show that the Ramsey fringes decay 
slower with increasing ion number $N$ since the rate $\Gamma_c$  
decreases with increasing $N$. In the cavity case, this would be equivalent 
to a single-atom cooperativity parameter in the superradiant laser that 
scales inversely with the number of atoms.

We observe that a Ramsey fringe decay rate of around $10 \Gamma_c$, 
achieved with Doppler cooling ($\bar{n}_{\COM} \approx 4.7)$, can be as 
low as $\sim \Gamma_c$ if the ions are cooled to their 
zero-point motion ($\bar{n}_{\COM} \approx 0)$. An ensemble of spins 
synchronized via this scheme can give fringes that decay slower 
than what the decay and dephasing processes dictate for uncorrelated spins.
This also relies on using a repump scheme that has a negligible 
branching ratio back to the initial state.

With this mapping, we can apply the unique tools that ion traps offer to 
study spin synchronization from steady-state superradiance. The ability to 
address single ions or specific subsets of ions in a trap can greatly 
advance studies of synchronization of two ensembles of ions that share the 
same damped normal mode 
\cite{[See ] [ for an analogous study in the cavity case.] xuPRL2014}. 
Ion traps could be used to explore quantum phase 
transitions between synchronized and unsynchronized phases, studying the 
build-up of correlations at the individual spin level. Recently, a 
cooling scheme for atoms in cavities that takes advantage of the collective 
interactions via a damped cavity mode has been proposed 
\cite{xuPRL2016}. It will be 
interesting to see if there are analogies to this `supercooling' in ion 
trap systems.

\begin{acknowledgments}
The authors acknowledge helpful discussions with Martin G\"{a}rttner, Yiheng 
Lin and Peiru He. This work is supported by the National Science Foundation 
under Grants PHY-1521080, PHY-1404263 and PHY-1125844, and by DARPA. This 
manuscript is the contribution of NIST and is not subject to US copyright.  
\end{acknowledgments}

\appendix

\section{\label{app:normal_mode_damping}Damping of the normal modes}

For brevity, we will use the notation $\rho \equiv \rho_{\tn}$ and 
$\mu \equiv \mu_{\neff}$ in this section.

In the Lamb-Dicke regime ($\langle (k z_m)^2 \rangle^{1/2} \ll 1$), we can 
expand the RHS of the master equation (\ref{eqn:tau_phonon_ME}) in powers of 
\{$\eta_n^\tau$\}. Up to second order in \{$\eta_n^\tau$\} we get

\begin{equation}
\dot{\rho} = (\lv_S + \lv_{R} + \lv_{SR}) \rho,
\label{eqn:tau_phonon_BM_start}
\end{equation}

\noindent where 

\begin{eqnarray}
&&\lv_S \rho = -i [ \sum_n \omega_n b_n^\dagger b_n, \rho], \nonumber\\*
&&\lv_R \rho = -i [ -\frac{1}{2} \Delta_\tau \sum_m \tau_m^z, \rho] 
+ \frac{\Gamma}{2} \sum_m \mathcal{D}[\tau_m^-] \rho, \; \text{and} 
\nonumber\\*
&&\lv_{SR} \rho = \lv_{SR}^{(1)} \rho + \lv_{SR}^{(2)} \rho, 
\label{eqn:tau_phonon_terms_pert}
\end{eqnarray}

\noindent with

\begin{eqnarray}
&&\lv_{SR}^{(1)} \rho = 
-i [ \frac{\Omega_\tau}{2} \sum_m (k z_m) (\tau_m^- + \tau_m^+), \rho], 
\;\text{and} \nonumber\\*
&&\lv_{SR}^{(2)} \rho = \frac{\Gamma_\tau}{2} \langle u^2 \rangle 
\sum_m \tau_m^-\big( 2 (k z_m) \rho (k z_m)  \nonumber\\*
&& \hphantom{\lv_{SR}^{(2)} \rho = 
\frac{\Gamma_\tau}{2} \langle u^2 \rangle \sum_m } 
- (k z_m)^2 \rho - \rho (k z_m)^2 \big) \tau_m^+.
\end{eqnarray}

Here $S$ denotes the system of normal modes, and $R$ denotes the 
reservoir of $\tau$ ions.

We first transform to an interaction picture with $\lv_0 = \lv_S + \lv_R$ 
\cite{[For a concise introduction to superoperators see ] 
carmichael2010statistical}. 
We then have

\begin{equation}
\dot{\td{\rho}} = \td{\lv}_{SR} \td{\rho} = (\td{\lv}_{SR}^{(1)} 
+ \td{\lv}_{SR}^{(2)}) \td{\rho},
\label{eqn:tau_phonon_int_pic} 
\end{equation}

where $\td{\rho} = e^{-(\lv_S+\lv_R)t} \rho$ and 
$\td{\lv}_{SR} = e^{-(\lv_S+\lv_R)t} \lv_{SR} e^{(\lv_S+\lv_R)t}$. 
Integrating Eq.~(\ref{eqn:tau_phonon_int_pic}) and substituting 
the formal solution of $\td{\rho}(t)$ back into Eq. 
(\ref{eqn:tau_phonon_int_pic}) gives (up to second order in 
\{$\eta_n^\tau$\}),

\begin{equation}
\dot{\td{\rho}} = \td{\lv}_{SR}^{(1)} \td{\rho}(0) 
+ \td{\lv}_{SR}^{(2)} \td{\rho}(0) 
+ \int_0^t dt^\prime \td{\lv}_{SR}^{(1)} (t) \td{\lv}_{SR}^{(1)} (t^\prime) 
\td{\rho}(t^\prime). 
\end{equation}  

When the couplings $\{\Omega_\tau \eta_n^\tau\} \ll \Gamma_\tau$, the $\tau$ 
ions serve as a reservoir of ions in a steady-state dictated by the 
reservoir Liouvillian $\lv_R$. In this case, the steady-state $R_0$ is the 
ground state of the $\tau$ ions, i.e., 
$R_0 = \ket{g}\bra{g}^{\otimes N_\tau}$. Starting from an initial 
uncorrelated state $\td{\rho}(0) = \td{\mu}(0) R_0$, we then use a 
decorrelation approximation to write $\td{\rho}(t) \approx \td{\mu}(t) R_0$ 
for subsequent times, and trace out the spin degrees of freedom of the $\tau$ 
ions:

\begin{eqnarray}
&&\dot{\td{\mu}} = \text{Tr}_R [\td{\lv}_{SR}^{(1)}(t) \td{\mu}(0) R_0] + 
\text{Tr}_R [\td{\lv}_{SR}^{(2)}(t) \td{\mu}(0) R_0] \nonumber\\*
&&\hphantom{\dot{\td{\mu}} =} 
+ \int_0^t dt^\prime \text{Tr}_R [\td{\lv}_{SR}^{(1)}(t) 
\td{\lv}_{SR}^{(1)}(t^\prime) \td{\mu}(t^\prime) R_0].
\end{eqnarray}

The first term vanishes because $\ev{\tau_m^\pm} = 0$ in the ground state, 
and the second term vanishes because $\ev{\tau_m^+ \tau_m^-}$ is zero 
in the ground state.

The structure of $\lv_{SR}^{(1)}$ (Eq.~(\ref{eqn:tau_phonon_terms_pert})) 
suggests that we need to find the time evolution of the superoperators 
$\td{\tau}_m^\pm \otimes I$ and $I \otimes (\td{\tau}_m^\mp)^T$.
 
This notation for a superoperator is to be understood as follows. 
Let $A,B$ be two operators acting on a Hilbert space spanned by $\ket{e}, 
\ket{g}$. Then the action of a superoperator $\lv = A \otimes (B)^T$ on a 
vector in the corresponding Liouville space, for eg. 
$\lvket{\lambda} = \ket{e} \bra{g}$ is $\lv \lvket{\lambda} = 
A\ket{e}\bra{g}B$. 

From 
$\td{\lv}_I = e^{-(\lv_S+\lv_R)t} \lv_{I} e^{(\lv_S+\lv_R)t}$, we have,

\begin{equation}
\dot{\td{\lv}}_I = [\td{\lv}_I, \lv_R]. 
\end{equation}   

This immediately gives the following complete set of equations:

\begin{eqnarray}
\frac{d}{dt} \td{\tau}_m^- \otimes I && = 
-(\frac{\Gamma_\tau}{2} -i \Delta_\tau) \td{\tau}_m^- \otimes I, \nonumber\\*
\frac{d}{dt} \td{\tau}_m^+ \otimes I && = 
(\frac{\Gamma_\tau}{2} -i \Delta_\tau) \td{\tau}_m^+ \otimes I + 
\Gamma_\tau \td{\tau}_m^z \otimes (\td{\tau}_m^+)^T, \nonumber\\*
\frac{d}{dt} \td{\tau}_m^z \otimes I && = 
-2 \Gamma_\tau \td{\tau}_m^- \otimes (\td{\tau}_m^+)^T.
\end{eqnarray}

The time evolution of the superoperators is then given by

\begin{eqnarray}
\td{\tau}_m^- \otimes I (t) && = 
\tau_m^- \otimes I e^{-(\frac{\Gamma_\tau}{2}-i\Delta_\tau)t},\nonumber\\*
\td{\tau}_m^+ \otimes I (t) && = 
\tau_m^+ \otimes I e^{(\frac{\Gamma_\tau}{2}-i\Delta_\tau)t} 
+ \tau_m^z \otimes (\tau_m^+)^T \times \nonumber\\*
&&\left(  e^{(\frac{\Gamma_\tau}{2}-i\Delta_\tau)t} 
- e^{-(\frac{\Gamma_\tau}{2}+i\Delta_\tau)t} \right).
\end{eqnarray}

Hermitian conjugation of the above two equations gives the time evolution of 
the other two superoperators. 

The master equation can then be written as 

\begin{eqnarray}
&&\dot{\td{\mu}} = \nonumber\\*
&& - \sum_m \sum_n \sum_k (\frac{\Omega_\tau}{2} \eta_n^\tau \mathcal{M}_{mn}) 
(\frac{\Omega_\tau}{2} \eta_k^\tau \mathcal{M}_{mk}) 
\times \int_0^t dt^\prime \biggl[  \nonumber\\*
&& \hphantom{+}  \{\td{b}_n(t) + \td{b}_n^\dagger(t)\} 
\{\td{b}_k(t^\prime) + \td{b}_k^\dagger(t^\prime)\} \td{\mu}(t^\prime) 
e^{-(\frac{\Gamma_\tau}{2}-i\Delta_\tau)(t-t^\prime)} \nonumber\\*
&&  - \{\td{b}_n(t) + \td{b}_n^\dagger(t)\} \td{\mu}(t^\prime) 
\{\td{b}_k(t^\prime) + \td{b}_k^\dagger(t^\prime)\} 
e^{-(\frac{\Gamma_\tau}{2}+i\Delta_\tau)(t-t^\prime)} \nonumber\\*
&& - \{\td{b}_k(t^\prime) + \td{b}_k^\dagger(t^\prime)\} \td{\mu}(t^\prime) 
\{\td{b}_n(t) + \td{b}_n^\dagger(t)\} 
e^{-(\frac{\Gamma_\tau}{2}-i\Delta_\tau)(t-t^\prime)} \nonumber\\*
&& + \td{\mu}(t^\prime) \{\td{b}_k(t^\prime) + \td{b}_k^\dagger(t^\prime)\} 
\{\td{b}_n(t) + \td{b}_n^\dagger(t)\} 
e^{-(\frac{\Gamma_\tau}{2}+i\Delta_\tau)(t-t^\prime)} \biggr]. \nonumber\\*
\label{eqn:tau_phonon_ME_time_evolution}
\end{eqnarray}

The time evolution of the mode annihilation and creation operators is given 
by $\td{b}_n(t) = b_n e^{-i\omega_n t}$ and 
$\td{b}_n^\dagger (t) = b_n^\dagger e^{i\omega_n t}$. 

If $\td{\mu}(t^\prime)$ doesn't change significantly on the timescale 
$\Gamma_\tau^{-1}$ for the decay of correlations in the reservoir of 
$\tau$-spins, we can perform a Markov approximation and set 
$\td{\mu}(t^\prime) \approx \td{\mu}(t)$ in Eq. 
(\ref{eqn:tau_phonon_ME_time_evolution}). This is reasonable since the 
damping rates of the normal modes are of the order of 
$\frac{(\Omega_\tau \eta_n^\tau)^2}{\Gamma_\tau}$, and for laser intensities 
such that $\Omega_\tau \eta_n^\tau \ll \Gamma_\tau$, this implies 
$\frac{(\Omega_\tau \eta_n^\tau)^2}{\Gamma_\tau} \ll \Gamma_\tau$. Further, 
we can extend the upper limit of the integration to $\infty$ since for 
significant evolution of $\mu(t)$, we are interested in 
$t \gg \Gamma_\tau^{-1}$.

After performing the integration over $\chi = t-t^\prime$ in Eq.~(\ref{eqn:tau_phonon_ME_time_evolution}), we encounter terms rotating with 
frequencies $\omega_n + \omega_k$ and $\omega_n-\omega_k$. The former terms 
are rapidly oscillating, and can be dropped. 

Performing the reverse transformation $\mu = e^{\lv_S t} \td{\mu}$ gives 
us the master equation for the damping of the normal modes, which accounts 
for the coupling between the modes as well:

\begin{eqnarray}
\frac{d}{dt}\mu = 
&&-i\left[\sum_n\omega_n^\prime b_n^\dagger b_n, \mu \right] \nonumber\\*
&& + \sum_n D_{n,n}^- (2 b_n \mu b_n^\dagger 
-b_n^\dagger b_n \mu - \mu b_n^\dagger b_n) \nonumber\\*
&& + \sum_n D_{n,n}^+ (2 b_n^\dagger \mu b_n 
- b_n b_n^\dagger \mu - \mu b_n b_n^\dagger) \nonumber\\*
&& -i \sum_{n \neq k} C_{k,n}^- (b_n \mu b_k^\dagger - b_k \mu b_n^\dagger 
+ b_n^\dagger b_k \mu - \mu b_k^\dagger b_n) \nonumber\\*
&& -i \sum_{n \neq k} C_{k,n}^+ (b_n^\dagger \mu b_k - b_k^\dagger \mu b_n 
+ b_n b_k^\dagger \mu - \mu b_k b_n^\dagger) \nonumber\\*
&&    \sum_{n \neq k} D_{k,n}^- (b_n \mu b_k^\dagger + b_k \mu b_n^\dagger 
- b_n^\dagger b_k \mu - \mu b_k^\dagger b_n) \nonumber\\*
&&    \sum_{n \neq k} D_{k,n}^+ (b_n^\dagger \mu b_k + b_k^\dagger \mu b_n 
- b_n b_k^\dagger \mu - \mu b_k b_n^\dagger). \nonumber\\*
\label{eqn:doppler_full}
\end{eqnarray}

Here the coefficients are given by

\begin{eqnarray}
&&\omega_n^\prime = \omega_n + R_{n,n}^- (\Delta_\tau + \omega_n) 
+ R_{n,n}^+ (\Delta_\tau - \omega_n), \nonumber\\*
&&C_{k,n}^\pm = R_{k,n}^\pm (\Delta_\tau \mp \omega_k), \nonumber\\* 
&&D_{k,n}^\pm = R_{k,n}^\pm \frac{\Gamma_\tau}{2}, \; \text{where} 
\nonumber\\*
&&R_{k,n}^\pm = \frac{ 
\sum_m (\frac{1}{2} \Omega_\tau \eta_n^\tau \mathcal{M}_{mn}) 
(\frac{1}{2} \Omega_\tau \eta_k^\tau \mathcal{M}_{mk}) } 
{\frac{\Gamma_\tau^2}{4} + (\Delta_\tau \mp \omega_n)^2 }.
\end{eqnarray}

The Doppler cooling introduces couplings between the different modes, with 
the coupling strengths between two modes decreasing as their frequency 
separation increases. The result 
of such mode cross-coupling is to introduce an admixture of other modes into 
the mode of interest, which in the example we consider is the highest 
frequency center-of-mass (COM) mode. The symmetric coupling of the COM mode 
to the ions then deterioriates, but the essential physics still remains the 
same. The situation is analogous to introducing a random component in the 
positions of atoms relative to the cavity standing wave in the superradiant 
laser. For simplicity, we assume these mode cross-couplings to be small and 
neglect them, use $D_n^\pm \equiv D_{n,n}^\pm, R_{n}^\pm \equiv R_{n,n}^\pm$ 
to simplify the notation,  and arrive at the master equation describing 
the damping of individual normal modes (Eq.~(\ref{eqn:tau_phonon_final_ME})).

\section{\label{app:schrieffer_wolff}Schrieffer-Wolff formalism 
for the three-level $\sigma$ ions}

The idea is to work in operator space, i.e. in the vector space $S$ spanned 
by the vectors $\ket{1}\bra{1}, \ket{1}\bra{2}, \ldots, \ket{3}\bra{3}$. The 
Liouvillian describing the dynamics can be written as the sum of a zeroth 
order Liouvillian $\lv_{0}$ and a perturbation $\mathcal{V}$. Based on the 
eigenvalues $\{\lambda_i\}$ of $\lv_0$, the space $S$ can be partitioned 
into a slow subspace, spanned by eigenvectors with eigenvalue $0$, and a 
complementary fast subspace spanned by eigenvectors with non-zero 
eigenvalue \cite{kesslerPRA2012}. If the left and right eigenvectors 
associated with an eigenvalue $\lambda_i$ are $\lvbra{l_i}$ and $\lvket{r_i}$ 
respectively, the projectors $P$ and $Q$ onto the slow and fast subspaces are 

\begin{eqnarray}
P && = \sum_{i:\{\lambda_i\}=0} \lvket{r_i}\lvbra{l_i}, \nonumber\\*
Q && = 1 - P = \sum_{i:\{\lambda_i\} \neq 0} \lvket{r_i}\lvbra{l_i}. 
\end{eqnarray}

Any superoperator $A:S \rightarrow S$ can now be represented as

\begin{equation}
A = \left( \begin{array}{cc}
            A^P & A^- \\
            A^+ & A^Q  
            \end{array} \right) 
  = \left( \begin{array}{cc}
            PAP & PAQ \\
            QAP & QAQ  
            \end{array} \right). 
\end{equation} 

The perturbation $\mathcal{V}$ in general couples the slow and fast 
subspaces. The Schrieffer-Wolff formalism provides a systematic, 
order-by-order procedure to find the effective Liouvillian 
$\lv^{\text{eff}}$ in the slow subspace that arises from this coupling. 
Explicitly, at the first three orders of perturbation theory,

\begin{eqnarray}
\lv_{1}^{\text{eff}} && = \pt^P \nonumber\\*
\lv_{2}^{\text{eff}} && = -\pt^- \lv_0^{-1} \pt^+ \nonumber\\*
\lv_{3}^{\text{eff}} && = \pt^- \lv_0^{-1} \pt^Q \lv_0^{-1} \pt^+ - 
\frac{1}{2} \{\pt^P, \pt^- \lv_0^{-2} \pt^+\}_+,
\label{eqn:L_eff_pert}
\end{eqnarray}
  
\noindent where $\{A,B\}_+ = AB + BA$.  

Table~\ref{tab:bases_S} gives the notation we adopt for the basis vectors in 
$S$. We split the superoperator appearing in Eq.~(\ref{eqn:sigma_Raman_ME}) 
into a zeroth order Liouvillian $\lv_0$ and a perturbation $\pt$. $\lv_0$ is 
already diagonal in the chosen basis. The third column of 
Table~\ref{tab:bases_S} gives the eigenvalues associated with $\lv_0$ for 
each of the basis. Then, the subspace spanned by the eigenvectors with 
eigenvalue $0$, i.e. $\{\lvket{A_1}, \lvket{A_3}, \lvket{A_5}, \lvket{A_7}, 
\lvket{A_9} \}$ is the slow subspace.

\begin{table}[!htb]
\caption{\label{tab:bases_S}Basis vectors in operator space $S$. 
The chosen zeroth order Liouvillian $\lv_0$ is diagonal in the above 
basis. The eigenvalues of $\lv_0$ are given in the third column.}
\centering
\begin{tabular}{ccc}
\toprule
Notation & Basis & Eigenvalue \\\hline
$\lvket{A_1}$ & $\ket{1}\bra{1}$ & $0$	 \\
$\lvket{A_2}$ & $\ket{1}\bra{2}$ & $-i \Delta_1$	 \\
$\lvket{A_3}$ & $\ket{1}\bra{3}$ & $0$ 	 \\
$\lvket{A_4}$ & $\ket{2}\bra{1}$ & $i \Delta_1$	 \\
$\lvket{A_5}$ & $\ket{2}\bra{2}$ & $0$	 \\
$\lvket{A_6}$ & $\ket{2}\bra{3}$ & $i \Delta_2$	 \\
$\lvket{A_7}$ & $\ket{3}\bra{1}$ & $0$	 \\
$\lvket{A_8}$ & $\ket{3}\bra{2}$ & $-i \Delta_2$	 \\
$\lvket{A_9}$ & $\ket{3}\bra{3}$ & $0$	 \\
\botrule
\end{tabular}
\end{table}

We also write the perturbation $\pt$ explicitly as a matrix acting on $S$. 
A better insight is obtained if we write vectors and matrices in the 
following order of basis vectors: $\lvket{A_1}, \lvket{A_3}, \ldots, 
\lvket{A_9}, \lvket{A_2}, \ldots, \lvket{A_8}$. 
In this representation, $\pt$ is given by

\begin{widetext}
\begin{eqnarray}
&& \pt = \left( \begin{array}{cc}
            \pt^P & \pt^- \\
            \pt^+ & \pt^Q  
            \end{array} \right) \nonumber\\*
&& = \left(
\begin{array}{ccccc|cccc}
0 & 	0 &	 \Gamma_1 &	 0 &	 0 &	i \frac{g_1}{2} &	 
-i \frac{g_1^*}{2} &	 0 &	 0 \\
0 & 	-i (\Delta_1-\Delta_2) & 	0 &	 0 &	 0 &	 
i \frac{g_2}{2} &	 0 &	 -i \frac{g_1^*}{2} &	 0 \\
0 & 0 & -(\Gamma_1+\Gamma_2) & 0 & 0 & -i \frac{g_1}{2} & 
i \frac{g_1^*}{2} & i \frac{g_2^*}{2} & -i \frac{g_2}{2} \\
0 & 0 & 0 & i(\Delta_1-\Delta_2) & 0 & 0 & -i \frac{g_2^*}{2} & 
0 & i \frac{g_1}{2} \\
0 & 0 & \Gamma_2 & 0 & 0 & 0 & 0 & -i \frac{g_2^*}{2} & 
i \frac{g_2}{2} \\ \hline
i \frac{g_1^*}{2} & i \frac{g_2^*}{2} & -i \frac{g_1^*}{2} & 0 & 0 & 
-\frac{\Gamma_1+\Gamma_2}{2} & 0 & 0 & 0 \\
-i \frac{g_1}{2} & 0 & i \frac{g_1}{2} & -i \frac{g_2}{2} & 0 & 0 & 
-\frac{\Gamma_1+\Gamma_2}{2} & 0 & 0 \\
0 & -i \frac{g_1}{2} & i \frac{g_2}{2} & 0 & -i \frac{g_2}{2} & 0 & 0 & 
-\frac{\Gamma_1+\Gamma_2}{2} & 0 \\
0 & 0 & -i \frac{g_2^*}{2} & i \frac{g_1^*}{2} & i \frac{g_2^*}{2} & 0 & 
0 & 0 & -\frac{\Gamma_1+\Gamma_2}{2} \\
\end{array}
\right).
\end{eqnarray}
\end{widetext}

The matrix has been partitioned to show the various blocks that make up the 
perturbation. Using Eq.~(\ref{eqn:L_eff_pert}), we calculate 
$\lv^\text{eff} = \lv_{1}^{\text{eff}} + \lv_{2}^{\text{eff}} + 
\lv_{3}^{\text{eff}}$, which is the effective Liouvillian in the slow 
subspace. In terms of the symbols defined in 
Table~\ref{tab:sigma_Raman_eff_symbols}, the effective Liouvillian 
$\lv^\text{eff}$, correct up to $O(\frac{|g_{1,2}|^2}{\Delta^2})$ is given by,

\begin{widetext}
\begin{equation}
\left(
\begin{array}{ccccc}
-\Gamma_{13}	&	
i\frac{\Omega_R}{2} + \frac{\Gamma_{1,\times}-\Gamma_{3,\times}}{2}	& 
\Gamma_1 - 2 \Gamma_{11} - \Gamma_{31}	&	
-i\frac{\Omega_R^*}{2} + \frac{\Gamma_{1,\times}^*-\Gamma_{3,\times}^*}{2} & 
\Gamma_{31}	\\[10pt]
i\frac{\Omega_R^*}{2} - \frac{\Gamma_{1,\times}^*+\Gamma_{3,\times}^*}{2} & 
-i\delta_R - \frac{\Gamma_{13}+\Gamma_{31}+\Gamma_{11}+\Gamma_{33}}{2}	& 
-\frac{\Gamma_{1,\times}^*+\Gamma_{3,\times}^*}{2}	&	0	& 
-i\frac{\Omega_R^*}{2} - \frac{\Gamma_{1,\times}^*+\Gamma_{3,\times}^*}{2} 
\\[10pt]
0	&	0	&	
\begin{aligned} -(\Gamma_1+\Gamma_2) +& 2\Gamma_{11}+\Gamma_{13} \\ 
+&\Gamma_{31}+2\Gamma_{33}\end{aligned}	& 
0	&	0	\\[10pt]
-i\frac{\Omega_R}{2} -\frac{\Gamma_{1,\times}+\Gamma_{3,\times}}{2} & 
0	&	-\frac{\Gamma_{1,\times}+\Gamma_{3,\times}}{2}	& 
i\delta_R -\frac{\Gamma_{13}+\Gamma_{31}+\Gamma_{11}+\Gamma_{33}}{2}	& 
i\frac{\Omega_R}{2}-\frac{\Gamma_{1,\times}+\Gamma_{3,\times}}{2}	
\\[10pt]
\Gamma_{13}	&	
-i\frac{\Omega_R}{2} -\frac{\Gamma_{1,\times}-\Gamma_{3,\times}}{2}	& 
\Gamma_2-\Gamma_{13}-2\Gamma_{33}	&	
i\frac{\Omega_R^*}{2} -\frac{\Gamma_{1,\times}^*-\Gamma_{3,\times}^*}{2} & 
-\Gamma_{31} \\
\end{array}
\right).
\end{equation} 
\end{widetext}

Here, the quantities $\Gamma_{1,\times} = 
\Gamma_1 \frac{g_1 g_2^*}{4\Delta^2}$ and $\Gamma_{3,\times} = 
\Gamma_2 \frac{g_1 g_2^*}{4\Delta^2}$.  The master equation for the $\sigma$ 
ion in the slow space is then $\dot{\mu} = \lv^\text{eff} \mu$, and this is 
given in Eq.~(\ref{eqn:sigma_Raman_eff_ME}) for a collection of $\sigma$ 
ions.

There are two points of note here. Firstly, if the system starts within 
the $\ket{1},\ket{3}$ manifold spanned by $\lvket{A_1}, \lvket{A_3}, 
\lvket{A_7} \text{and} \lvket{A_9}$, then it stays within that manifold. 
Then we do not need to consider the $\lvket{A_5}$ state. Secondly, 
the terms proportional to $\Gamma_{1,\times}$ and $\Gamma_{3,\times}$ 
give rise to certain cross-terms. A typical cross-term in the master 
equation appears as 

\begin{equation}
\frac{\Gamma_{1,\times}}{2} (-\sigma^+\mu - \sigma^z\mu\sigma^+).  
\label{eqn:cross_term}
\end{equation}   

We intend to couple the effective two-level system formed by the $\sigma$ 
ions to their external motion by tuning the Raman lasers to the red 
vibrational sideband. In that case, the only significant contributions to 
ion $l$ from a term such as (\ref{eqn:cross_term}) will be of the approximate 
form

\begin{equation}
\frac{1}{2} \frac{\Gamma_1}{\Delta} \sum_n \mathcal{F}_{ln} \sigma_l^+ b_n\mu.
\end{equation} 

The spin-motion coupling strength in this term is a factor of 
$\frac{\Gamma}{\Delta}$ smaller than the coherent spin-motion coupling 
present in the Hamiltonian terms. Subsequently, when we treat the 
spin-motion coupling perturbatively in comparison with the damping of 
the normal modes, the contribution from these cross-terms will be 
$\frac{\Gamma^2}{\Delta^2}$ smaller than the contribution from the 
Hamiltonian terms. Hence we neglect these cross-terms while writing down 
Eq.~(\ref{eqn:sigma_Raman_eff_ME}).

\section{\label{app:eff_spin_spin_model}Effective spin-spin model: 
Interaction of $\sigma$-spins with damped normal modes}

We will use the notation $\mu \equiv \mu_\sigsub$ in this section. 
Starting with Eq.~(\ref{eqn:sigma_phonon_BM_start}), we first transform 
to an interaction picture with $\lv_0 = \lv_R$. Following the steps outlined 
in the beginning of Appendix~\ref{app:normal_mode_damping}, we arrive at the 
following integro-differential equation for the reduced density matrix 
$\mu(t)$ describing the $\sigma$-spins only:

\begin{eqnarray}
\dot{\mu}(t) =  &&\text{Tr}_R [\td{\lv}_{SR}(t) \mu(0) R_0] \nonumber\\*
&& + \int_0^t dt^\prime \text{Tr}_R [\td{\lv}_{SR}(t) \td{\lv}_{SR}(t^\prime) 
\mu(t^\prime) R_0].
\label{eqn:sigma_phonon_Born_ME}
\end{eqnarray} 

Here, we have used a decorrelation approximation to write 
$\td{\rho}(t) \approx \mu(t) R_0$, where $R_0$ is the steady-state 
density matrix for the normal modes under the action of $\lv_R$. Once again, 
we start from an initial uncorrelated state: $\rho(0) = \mu(0) R_0$. 
Note that the density matrix $\mu(t)$ and the $\sigma$-spin operators 
do not have overhead tilde ($\sim$) in this Appendix 
since $\lv_S = 0$ in the present case.

Under the action of $\lv_R$, the steady-state density matrices of each of 
the normal modes are thermal states. The first term on the RHS of Eq. 
(\ref{eqn:sigma_phonon_Born_ME}) vanishes, since the expectation values 
$\ev{b_n}, \ev{b_n^\dagger}$ are zero in a thermal state. In order to 
evaluate the second term, we need to find the time evolution of the 
superoperators $\td{b}_n \otimes I, \td{b}_n^\dagger \otimes I, 
I \otimes (\td{b}_n)^T \text{and} \; I \otimes (\td{b}_n^\dagger)^T$. 
Following the lines of the procedure we adopted in finding the time 
evolution of the superoperators $\td{\tau}_m^- \otimes I$, etc. in 
Appendix~\ref{app:normal_mode_damping}, we get,

\begin{eqnarray}
&&\frac{d}{dt} \td{b}_n \otimes I (t) = -\left(\frac{\kappa_n}{2} (1+2\nbar_n) 
+ i \td{\delta}_n\right) \td{b}_n \otimes I \nonumber\\* 
&&\hphantom{\frac{d}{dt} \td{b}_n \otimes I (t) =} 
+ \kappa_n \nbar_n I \otimes (\td{b}_n)^T, \nonumber\\*
&&\frac{d}{dt} I \otimes (\td{b}_n)^T = \left(\frac{\kappa_n}{2}(1+2\nbar_n) 
- i \td{\delta}_n\right) I \otimes (\td{b}_n)^T \nonumber\\*
&&\hphantom{\frac{d}{dt} I \otimes (\td{b}_n)^T =} 
- \kappa_n (1+\nbar_n) \td{b}_n \otimes I.
\end{eqnarray}

Solving the above pair of coupled differential equations, we get,

\begin{eqnarray}
&&\td{b}_n \otimes I (t) = \nonumber\\*
&&\hphantom{\td{b}_n \otimes I }\nbar_n (I \otimes (b_n)^T - b_n \otimes I) 
e^{\left( \frac{\kappa_n}{2} -i \td{\delta}_n\right)t} \nonumber\\*
&&\hphantom{\td{b}_n \otimes I } 
+ ((1+\nbar_n) b_n \otimes I - \nbar_n I \otimes (b_n)^T) 
e^{-\left( \frac{\kappa_n}{2} + i\td{\delta}_n\right)t}, \nonumber\\*
&&I \otimes (\td{b}_n)^T (t) = \nonumber\\* 
&&\hphantom{I \otimes (\td{b}_n)^T } 
(1+\nbar_n) ( I \otimes (b_n)^T - b_n \otimes I) 
e^{\left( \frac{\kappa_n}{2} -i\td{\delta}_n\right)t} \nonumber\\*
&&\hphantom{I \otimes (\td{b}_n)^T } 
+ ( (1+\nbar_n) b_n \otimes I - \nbar_n I \otimes (b_n)^T ) 
e^{-\left( \frac{\kappa_n}{2} + i\td{\delta}_n\right)t}. \nonumber\\* 
\end{eqnarray}

Hermitian conjugation of the above two equations gives the time evolution of 
$ I \otimes (\td{b}_n^\dagger)^T \text{and} \; \td{b}_n^\dagger \otimes I$. 
Using the fact that $\ev{b_n^\dagger b_n} = \nbar_n$, we can now perform the 
trace over the reservoir of normal modes in Eq. 
(\ref{eqn:sigma_phonon_Born_ME}) to arrive at an expression involving 
intergrals over the $\sigma$-spin operators, $\mu(t^\prime)$ and 
complex exponentials. As an example, we consider one of the terms that occur 
in this expression:

\begin{equation}
- \sum_{l,m,n} \mathcal{F}_{ln} \mathcal{F}_{mn}^* (1+\bar{n}_n) 
\int_0^t dt^\prime \sigma_l^+ \sigma_m^-  \mu(t^\prime) 
e^{-\left( \frac{\kappa_n}{2} + i \td{\delta}_n \right) (t-t^\prime)}.
\label{eqn:eff_spin_spin_example_term}
\end{equation}    

We perform a Markov approximation by setting 
$\mu(t^\prime) \approx \mu(t)$. For significant evolution of $\mu(t)$, 
we are interested in evolution over times that are large compared to the 
timescales of the reservoir correlations. Only the upper limit of integration 
in terms like \ref{eqn:eff_spin_spin_example_term} contribute in this 
coarse-graining procedure. 

We then evaluate the simple time integrals over 
complex exponentials and group the coherent and dissipative parts 
separately. We then account for the incoherent Raman processes and the 
incoherent repumping, and arrive at the effective spin-spin model described 
by the master equation (\ref{eqn:sigma_sigma_eff_ME}), which is the starting 
point for our numerical analysis. 

\emph{A note on the validity of approximations:} To stop at second-order in 
perturbation theory, the timescale for the system-reservoir 
interaction must be long compared to the reservoir correlation time 
\cite{cohen1992atom}. Further, the Markov approximation requires that 
the timescale for significant evolution of the system $T_S$ is long 
compared to the reservoir correlation time. In a minimal model 
where the spin-spin interactions are mediated only by the COM mode 
and the other modes are neglected, the perturbation strength and 
fastest timescale for the $\sigma$-spins are determined by the 
collectively-enhanced spontaneous emission rate, given 
by $N_\sigma \Gamma_{\COM} (1 + \nbar_{\COM})$, 
where $\Gamma_{\COM} = \mathcal{F}^2_{\COM}/\kappa_{\COM}$. Since 
the correlation time for the COM mode is set by $\kappa_{\COM}$, 
we require 
$\kappa_{\COM} \gg N_\sigma \Gamma_{\COM} (1 + \nbar_{\COM}) \gtrsim T_S$ 
for second-order perturbation theory and the Markov approximation to be valid.

\section{\label{app:c_number_langevin}
Numerical simulation using $c$-number Langevin equations}

We start by writing the quantum Langevin equations (QLE) for the spin 
operators $\sigma_i^x$, $\sigma_i^y$ and $\sigma_i^z$ for a spin $i$ from 
the master equation (\ref{eqn:sigma_sigma_eff_ME}):

\begin{eqnarray}
\allowdisplaybreaks
\frac{d}{dt} \sigma_i^x &&= D_i^x + F_i^x \nonumber\\
&&= -\left\{\Gamma_{ii}^- + \Gamma_{ii}^+ + \frac{\Gamma_{31}}{2} 
+ \frac{\Gamma_{13}+w}{2} + \frac{\Gamma_d}{2}\right\} \sigma_i^x \nonumber\\
&&\hphantom{= } - B_i \sigma_i^y + \sum_{j \neq i} 
(\Gamma_{ji}^- - \Gamma_{ji}^+) \sigma_i^z \sigma_j^x \nonumber\\
&&\hphantom{= } + \sum_{j \neq i} J_{ji} \sigma_i^z \sigma_j^y + F_i^x, 
\nonumber\\
\frac{d}{dt} \sigma_i^y &&= D_i^y + F_i^y \nonumber\\
&&= -\left\{\Gamma_{ii}^- + \Gamma_{ii}^+ + \frac{\Gamma_{31}}{2} 
+ \frac{\Gamma_{13}+w}{2} + \frac{\Gamma_d}{2}\right\} \sigma_i^y \nonumber\\
&&\hphantom{= } + B_i \sigma_i^x + \sum_{j \neq i} 
(\Gamma_{ji}^- - \Gamma_{ji}^+) \sigma_i^z \sigma_j^y \nonumber\\
&&\hphantom{= } - \sum_{j \neq i} J_{ji} \sigma_i^z \sigma_j^x + F_i^y, 
\nonumber\\
\frac{d}{dt} \sigma_i^z &&= D_i^z + F_i^z \nonumber\\
&&= -\left\{ 2 ( \Gamma_{ii}^- + \Gamma_{ii}^+ ) + \Gamma_{31} 
+ \Gamma_{13} + w \right\} \sigma_i^z  \nonumber\\
&&\hphantom{= }+ \left\{ \Gamma_{13}+w - 
\left(2(\Gamma_{ii}^- - \Gamma_{ii}^+) + \Gamma_{31}\right) \right\} 
\nonumber\\
&&\hphantom{= } - \sum_{j \neq i} (\Gamma_{ji}^- - \Gamma_{ji}^+) 
(\sigma_i^x \sigma_j^x + \sigma_i^y \sigma_j^y) \nonumber\\
&&\hphantom{= } - \sum_{j \neq i} J_{ji} 
(\sigma_i^x \sigma_j^y - \sigma_i^y \sigma_j^x) + F_i^z. 
\label{eqn:spin_qle}
\end{eqnarray}

Here, $F_i^x$, $F_i^y$ and $F_i^z$ are operators that account for the noise 
because of coupling to an external environment. These noise operators are 
correlated according to

\begin{equation}
\langle F_i^\mu (t) F_j^\nu (t') \rangle = 
2 \langle D_{ij}^{\mu \nu} \rangle \delta(t-t^\prime),
\end{equation} 

where $\mu, \nu = x,y,z$ and $i,j$ are the spin indices. The generalized 
Einstein relation \cite{meystre1998elements} can be used to determine the 
correlation matrix elements 
$2 \langle D_{ij}^{\mu \nu} \rangle$:

\begin{equation}
2 \langle D_{ij}^{\mu \nu} \rangle = 
- \langle \sigma_\mu^i D_\nu^j \rangle 
- \langle D_\mu^i \sigma_\nu^j \rangle 
+ \frac{d}{dt} \langle \sigma_\mu^i \sigma_\mu^j \rangle.
\end{equation} 

Next, we perform a quantum-classical correspondence by associating a 
$c$-number with each of the spin operators, i.e. 
$\sigma_i^x \leftrightarrow s_i^x$, $\sigma_i^y \leftrightarrow s_i^y$ 
and $\sigma_i^z \leftrightarrow s_i^z$. The equations of motion for these 
$c$-numbers are obtained from the QLEs~(\ref{eqn:spin_qle}) by replacing the 
quantum operators with their corresponding $c$-numbers. The quantum 
noise operators $F_i^\mu$ are replaced by $c$-number noise terms 
$\mathcal{F}_i^\mu$.

We use symmetric correspondence to match the correlations of the $c$-number 
noise terms $\mathcal{F}_i^\mu$ with the correlations of the quantum 
noise operators $F_i^\mu$, i.e.

\begin{eqnarray}
&& \langle \mathcal{F}_i^\mu (t) \mathcal{F}_j^\nu (t') \rangle =  
2 \mathcal{D}_{ij}^{\mu \nu} \delta(t-t^\prime) \; \text{with} \nonumber\\*
&& 2 \mathcal{D}_{ij}^{\mu \nu} = D_{ij}^{\mu \nu} + D_{ji}^{\nu \mu}.
\end{eqnarray}

The elements of the correlation matrix $2 \mathcal{D}_{ij}^{\mu \nu}$ are 
summarized in Eq.~(\ref{eqn:cnum_diff_mat_elts}).

\begin{eqnarray}
&& 2 \mathcal{D}_{ii}^{x x} = 2 \mathcal{D}_{ii}^{y y} =  
2 (\Gamma_{ii}^- + \Gamma_{ii}^+) + \Gamma_{31} + (\Gamma_{13}+w) + \Gamma_d , 
\nonumber\\*
&& 2 \mathcal{D}_{ii}^{x y} = 0 , \nonumber\\*
&& 2 \mathcal{D}_{ii}^{z z} = 2 \left( w + \Gamma_{13} + \Gamma_{31} 
+ 2 (\Gamma_{ii}^- + \Gamma_{ii}^+) \right) \nonumber\\*
&&\hphantom{2 \mathcal{D}_{ii}^{z z} =} 
+ 2 \left(  \Gamma_{31} + 2 (\Gamma_{ii}^- - \Gamma_{ii}^+)  
- (w + \Gamma_{13})  \right) \langle \sigma_i^z \rangle , \nonumber\\*
&& 2 \mathcal{D}_{ii}^{x z} = \left( \Gamma_{31} 
+ 2 (\Gamma_{ii}^- - \Gamma_{ii}^+)  - (w + \Gamma_{13})  \right) 
\langle \sigma_i^x \rangle ,\nonumber\\*
&& 2 \mathcal{D}_{ii}^{y z} = \left( \Gamma_{31} 
+ 2 (\Gamma_{ii}^- - \Gamma_{ii}^+)  - (w + \Gamma_{13})  \right) 
\langle \sigma_i^y \rangle , \nonumber\\*
&& 2 \mathcal{D}_{ij}^{x x} = 2 \mathcal{D}_{ij}^{y y} = 
2 (\Gamma_{ij}^- + \Gamma_{ij}^+) \langle \sigma_i^z \sigma_j^z \rangle , 
\nonumber\\*
&& 2 \mathcal{D}_{ij}^{x y} = 0 , \nonumber\\*
&& 2 \mathcal{D}_{ij}^{z z} = 2 (\Gamma_{ij}^- + \Gamma_{ij}^+) 
\left( \langle \sigma_i^x \sigma_j^x \rangle 
+ \langle \sigma_i^y \sigma_j^y \rangle \right) , \nonumber\\* 
&& 2 \mathcal{D}_{ij}^{x z} = -2 (\Gamma_{ij}^- + \Gamma_{ij}^+) 
\langle \sigma_i^z \sigma_j^x \rangle , \nonumber\\*
&& 2 \mathcal{D}_{ij}^{y z} = -2 (\Gamma_{ij}^- + \Gamma_{ij}^+) 
\langle \sigma_i^z \sigma_j^y \rangle .   
\label{eqn:cnum_diff_mat_elts}
\end{eqnarray}

By construction, the diffusion matrix is symmetric, and this property can be 
used to obtain the other elements. We simulate the $3 N_\sigma$ $c$-number 
Langevin equations subject to the noise 
correlation matrix $2 \mathcal{D}$ with elements given by Eq. 
(\ref{eqn:cnum_diff_mat_elts}). Using vector notation, these stochastic 
differential equations (SDEs) can be written as

\begin{equation}
\frac{d}{dt} \vec{s} (t) = \vec{f}\left\{\vec{s}(t)\right\} + B(t) d\vec{W},
\end{equation}

\noindent where the $\{dW_j\}$ are independent gaussian random variables 
with zero mean and variance $dt$. The function $\vec{f}$ accounts for the 
drift part of the SDEs, while the matrix $B(t)$ is given by

\begin{equation}
\begin{aligned}
& B = \sqrt{2 \mathcal{D}} = V \sqrt{\Lambda} V^{-1}, \quad \text{where}\\
& 2 \mathcal{D} = V \Lambda V^{-1}
\end{aligned}
\end{equation} 

\noindent is the transformation that diagonalizes $2 \mathcal{D}$ to the 
diagonal matrix $\Lambda$. We use an explicit second order weak scheme 
\cite{kloeden2011numerical} to numerically integrate these SDEs.

\bibliography{aps_template_ion_trap_superradiance_paper}

\end{document}